\documentclass[journal,twoside]{IEEEtran}

\usepackage{cite}

\usepackage{amsmath,amssymb,amsfonts,amsthm}
\usepackage{algorithmic}
\usepackage{graphicx}
\usepackage{textcomp}
\usepackage{xcolor}
\newtheorem{thm}{Theorem}
\newtheorem{lem}[thm]{Lemma}
\newtheorem{prop}{Proposition}

\newtheorem{rem}{Remark}
\newtheorem{ass}{Assumption}
\allowdisplaybreaks

\usepackage{enumitem}
\setlist[enumerate]{leftmargin=*}
\setlist[itemize]{leftmargin=*}

\newcommand{\real}[0]{\mathbb R}

\DeclareSymbolFont{bbold}{U}{bbold}{m}{n}
\DeclareSymbolFontAlphabet{\mathbbold}{bbold}
\DeclareMathOperator*{\argmin}{argmin}
\DeclareMathOperator*{\spec}{spec}

\newcommand{\eucd}{\mathbb{R}}

\newcommand{\norm}[1]{\left\| #1 \right\|}
\newcommand{\T}{\mathsf{T}}

\def\BibTeX{{\rm B\kern-.05em{\sc i\kern-.025em b}\kern-.08em
    T\kern-.1667em\lower.7ex\hbox{E}\kern-.125emX}}

\begin{document}
\title{Closed-Loop Motion Planning for Differentially Flat Systems: A Time-Varying Optimization Framework}
\author{Tianqi Zheng, John W. Simpson-Porco, \IEEEmembership{Senior Member, IEEE}, and Enrique Mallada, \IEEEmembership{Senior Member, IEEE}
\thanks{T.~Zheng and E.~Mallada are with the Department
of Electrical and Computer Engineering, Johns Hopkins University. E-mail: \{tzheng8,mallada\}@jhu.edu. }
\thanks{J.~W.~Simpson-Porco is with the Department of Electrical and Computer Engineering, University of Toronto. Email: jw.simpson@utoronto.ca. }
\thanks{This work was supported by the NSF Global Centers program under Grant No.~2330450 and by the DOE Office of Science (ASCR) under Award No.~826565.}
}

\maketitle

\begin{abstract}\color{black}
Motion planning and control are two core components of the robotic autonomy stack. The standard way to combine them uses an offline/open-loop stage, \textit{planning}, which designs a feasible and safe trajectory, and an online/closed-loop stage, \textit{tracking}, which corrects for unmodeled dynamics and disturbances. This separation introduces conservatism in planning that becomes difficult to overcome as model complexity increases and real-time decisions must be made in a changing environment. This work addresses these challenges for differentially flat nonlinear systems by integrating planning and control into a single closed-loop task. We develop an optimization-based framework that steers a flat system to a trajectory implicitly defined via a constrained \textit{time-varying optimization} problem. Generalizing feedback linearization, our controllers effectively transform the flat system into an algorithm that seeks the optimal solution of the time-varying problem. Under sufficient regularity assumptions, we prove global asymptotic convergence to the time-varying minimizer. We illustrate the method on multi-robot tracking and robot obstacle avoidance problems.
\end{abstract}


\section{Introduction} \label{sec:introduction}

\IEEEPARstart{A}{utonomous} tasks in robotics and transportation involve sensing, localization, planning, and control~\cite{10.1177/02783649922067753,gonzalez2015review,elbanhawi2014sampling}. A particularly challenging step is motion planning~\cite{10.1177/02783649922067753,gonzalez2015review,elbanhawi2014sampling,kavraki1996probabilistic}, wherein an agent with uncertain information about its position and environment must devise an admissible, collision-free trajectory toward a destination, a task that has received widespread interest~\cite{dijkstra2022note, hart1968formal,stentz1993optimal,elbanhawi2014sampling,lavalle1998rapidly,kavraki1996probabilistic,lewis2012optimal,camacho2013model,zucker2013chomp,gonzalez2015review}, as it balances computational complexity and optimality while respecting the agent's dynamic capabilities.

Standard approaches to solving motion planning problems can be broadly categorized into three groups: Grid-based search (GBS), Sampling-based Planning (SBP), and Optimization-based (OB). GBS algorithms  assign each configuration of the dynamical system to a grid point and use graph search algorithms such as Dijkstra \cite{dijkstra2022note}, $A^*$ \cite{hart1968formal}, and $D^*$ \cite{stentz1993optimal} to find a path. 
GBS algorithms are simple to implement but scale poorly with the degrees of freedom of the configuration space \cite{lindemann2005current} and do not ensure dynamic feasibility of the path. %
SBP algorithms \cite{elbanhawi2014sampling}, such as rapidly-exploring random trees (RRTs) \cite{lavalle1998rapidly}, probabilistic roadmap methods (PRMs) \cite{kavraki1996probabilistic}, and their variants scale better for high-dimensional problems. However, optimality guarantees are usually absent, and path feasibility is only achieved via sufficiently dense sampling of either the configuration or action space \cite{lindemann2005current}.
On the other hand, OB algorithms such as 
direct multiple-shooting~\cite{bock1984multiple} and direct collocation \cite{biegler1984solution} explicitly consider the dynamic constraints in the optimization problems, providing by construction dynamically feasible trajectories which can be enforced to avoid collisions~\cite{zhang2020optimization,guthrie2022closed}. 
However, OB algorithms suffer from high computational costs, typically requiring solving a nonlinear programming problem without convergence guarantees~\cite{zucker2013chomp}. 

Two common features of the above-mentioned solutions are (a) the struggle between the computational complexity of the planning process and the need to enforce dynamic constraints and (b) the open-loop nature of the solution that does not account for unmodeled dynamics or disturbances. Thus, such methodologies are commonly complemented with a motion execution stage that implements a feedback controller that tracks the open-loop trajectory. {In practice, the planning stage is then re-executed either periodically, in a receding-horizon fashion, or upon events such as a detected infeasibility, at a rate that is application specific; between re-planning instants, the reference is followed in open loop.} Such an approach introduces conservatism in the planning stage to avoid collisions~\cite{10.1109/icra.2017.7989693,g2022phd-thesis}, which, when explicitly accounted for via robust methods, further increases the computational cost. This work explores an alternative: integrating planning and control as a single task, ensuring dynamic feasibility while accounting for changing conditions in real time. 

\smallskip

\subsubsection*{Contributions}
This work uses time-varying optimization (TV-O) to combine safe motion planning and control into a single \emph{closed-loop} task. We seek to encode planning goals and safety constraints as a TV constrained optimization problem and develop a general methodology to design closed-loop feedback controllers by drawing insights from mathematical optimization. Our methodology combines tools from differential flatness and optimization theory to develop controllers which effectively transform a dynamical system into an optimization algorithm that seeks to track the optimal solution of the aforementioned TV-O problem.

\begin{figure}[t]
\centerline{\includegraphics[width=0.92\columnwidth]{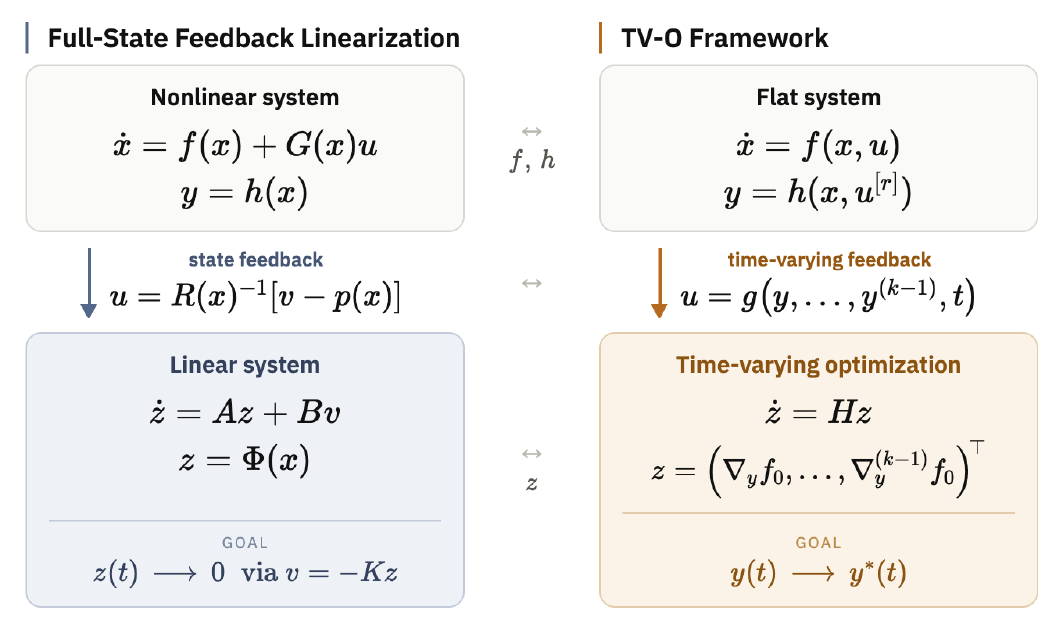}}
   \caption{{{The TV-O framework (right) generalizes full-state feedback linearization (left). The left panel assumes a control-affine system with total relative degree $n$, nonsingular $R(x)$, and a valid state transformation on the domain considered. Choosing $v=-Kz$ stabilizes the linearized system when $A-BK$ is Hurwitz. The right panel assigns stable linear dynamics ($H$ Hurwitz) to the objective's gradient and its time derivatives, yielding convergence to $y^*(t)$ under the stated assumptions.}}  }
\label{fig:framework}
\end{figure}

The contributions of our work are as follows:
\begin{itemize}\itemsep=2pt
    \item \emph{Planning and Control as TV-O.}
    We formulate a framework to encode planning and control goals within a TV-O problem, wherein planning goals are implicitly encoded as the (apriori unknown) optimal solution $y^*(t)$. This formulation allows us, in turn, to recast the control design problem as the problem of choosing an optimization algorithm.
    
    \item \emph{Flat Systems as Optimization Algorithms.}
    We provide a general methodology to design control laws that steer the output $y(t)$ of a differentially flat nonlinear system towards $y^*(t)$. Inspired by feedback linearization \cite{zsm2020acc,isidori2013nonlinear} (Fig. \ref{fig:framework}, left), the proposed methodology transforms any flat system of order $k$ into a time-varying optimization algorithm that depends on the first $k-1$ time derivatives of the objective function's gradient (Fig. \ref{fig:framework}, right).
    \item \emph{Theoretical Guarantees.} 
    {For differentially flat systems, under smoothness, strong convexity, and {mild regularity assumptions specified for each formulation (Assumptions~\ref{ass:MFCQ}--\ref{ass:lipshictz_continuity})}, we prove that the flat output $y(t)$ converges asymptotically to the (a priori unknown) optimizer $y^*(t)$; flatness is what enables this guarantee, as it lets the controller assign the full optimization dynamics to the flat output.}

    \item \emph{Extensions for Formation and Collision Avoidance.}
    We further extend our time-varying feedback optimization framework to allow for (i) the asymptotic satisfaction of time-varying equality constraints, which allow for specification of formation constraints, and (ii) hard inequality constraints, which can model collision avoidance specifications.

\end{itemize}

A preliminary version of this work appeared in \cite{zsm2020acc}. The present paper extends the work of \cite{zsm2020acc} in many ways, including extensions from feedback linearization to general flat systems, as well as the inclusion of equality and inequality constraints. The technical results of Sections \ref{sec:Unconstrained_TV_optimization_Framework} , \ref{sec:Equality Constrained TVO Framework} and \ref{sec:Ineq Constrained TVO Framework}
are new, as well as the numerical validations of Section \ref{sec:Numerical}.

\smallskip

\subsubsection*{Other related literature}
Our work broadly aligns with and contributes to the growing literature of \emph{online optimization with feedback loops}, \emph{network systems},  and \emph{algorithm design for time-varying optimization}.

\noindent\textbf{Online optimization with feedback loops} seeks to design online optimization algorithms to regulate the output of a system towards the optimal solution of an optimization problem. For the case of LTI systems, numerous works have designed controllers that track the optimal solution of (i) a static optimization problem~\cite{nelson2018integral, Lawrence2020Linear, menta2018stability}, and (ii) time-varying convex optimization problems \cite{colombino2019online}, including also input-output constraints~\cite{bianchin2021time}. Nonlinear system dynamics are considered for steering a physical system to a steady state that solves a predefined constrained static optimization problem~\cite{hauswirth2020timescale} and an unconstrained time-varying optimization problem~\cite{zsm2020acc}.

{\noindent\textbf{Feedback optimization}, a prominent branch of the line of work above, interconnects an optimization algorithm with a (pre-stabilized) plant so that the closed-loop \emph{equilibrium} is the optimizer, typically via first-order updates and timescale separation; recent applications include distributed and perception-based robotic coordination \cite{terpin2022distributed,carnevale2024nonconvex,cothren2022online}.}
{Our framework instead tracks the \emph{moving} optimizer of a time-varying problem with vanishing error, through the full nonlinear dynamics and without pre-stabilization, by exploiting differential flatness.}
\noindent\textbf{Online optimization of network systems} extends the above to distributed settings, regulating a network of agents to the minimizer of a convex problem \cite{Shahrapour2018Distributed,lin2016distributed}, including time-varying versions with inequality constraints \cite{sun2020distributed}, double-integrator dynamics \cite{rahili2016distributed}, and strict-feedback nonlinear dynamics \cite{huang2019distributed}. {\color{black} Extending our framework for this setting is a natural next step.}

\noindent\textbf{Time-varying optimization} provides a computationally frugal framework for online decision making when data arrive as streams and decisions must be made at high frequency \cite{simonetto2020time,rahili2016distributed,fazlyab2017prediction,simonetto2016class,simonetto2017prediction}. Online solvers have been proposed in both continuous \cite{rahili2016distributed,fazlyab2017prediction} and discrete time \cite{simonetto2016class,simonetto2017prediction}. Our work extends this literature by devising controllers for nonlinear differentially flat systems that track the optimizer of constrained time-varying problems.

\subsubsection*{Outline of the paper}
Section \ref{sec:preliminaries} reviews differential flatness. Section \ref{sec:ProblemStatement} states the problem, our assumptions, and two motivating examples. Sections \ref{sec:Unconstrained_TV_optimization_Framework}, \ref{sec:Equality Constrained TVO Framework}, and \ref{sec:Ineq Constrained TVO Framework} develop the framework for unconstrained, equality-constrained, and inequality-constrained problems, respectively. Section \ref{sec:Numerical} presents numerical experiments, and Section \ref{sec:Conclusion} concludes.

\subsubsection*{Notation} 
We use bold symbols to represent vectors ($\mathbf{x}$) and matrices ($\mathbf{A}$). {For $n \in \mathbb{Z}_{\geq 0}$, let $[n] := \{1, \dots, n\}$, with $[0] := \emptyset$.}   A square symmetric matrix $\mathbf{A}$ is positive (semi)definite, written as $\mathbf{A} \succ 0$ ($\mathbf{A} \succeq 0$), if and only if all the eigenvalues of $\mathbf{A}$ are positive (nonnegative). We further write $\mathbf{A}\succ \mathbf{B}$ ($\mathbf{A}\succeq \mathbf{B}$) whenever $\mathbf{A-B}\succ0$ ($\mathbf{A-B}\succeq0$). The symbol $ \otimes$  denotes the Kronecker product between two matrices. The Euclidean norm of a vector $\mathbf{x}$ is denoted by $\|\mathbf{x} \|_2$, and the spectral norm of a matrix $\mathbf{A}$ by $ \| \mathbf{A} \|_2$. The $k$-th order  derivative of $\mathbf{x}$ with respect to time $t$ is denoted as  $\mathbf{x}^{(k)}$. For a nonnegative integer $k$ we use the short-hand notation $ \mathbf{x}^{[k]} = (\mathbf{x},\mathbf{x}^{(1)}, \dots,\mathbf{x}^{(k)}  )$.

Given a sufficiently differentiable function  $f(\mathbf{x},t)$ of state $\mathbf{x} \in \mathbb{R}^n$ and time $t \in \real$, the gradient with respect to $\mathbf{x}$ (resp. $t$) is denoted by $\nabla_\mathbf{x} f(\mathbf{x},t)$  (resp. $\nabla_t f(\mathbf{x},t)$). When $x(t)$ itself depends on time, the total derivative (resp. $n$-th total derivative) of $\nabla_\mathbf{x} f(\mathbf{x}(t),t)$ with respect to $t$ is denoted by $\dot{\nabla}_\mathbf{x} f(\mathbf{x},t):=\frac{\mathrm{d}}{\mathrm{d}t}{\nabla}_\mathbf{x} f(\mathbf{x}(t),t)$ (resp. $\nabla^{(n)}_\mathbf{x} f(\mathbf{x},t)$).
The partial derivatives of $\nabla_\mathbf{x} f(\mathbf{x},t)$ with respect to $\mathbf{x}$ and $t$ are denoted by $\nabla_{\mathbf{xx}}f(\mathbf{x},t) := \frac{\partial}{\partial \mathbf{x}} \nabla_\mathbf{x} f(\mathbf{x},t) \in \eucd^{n\times n}$ and $\nabla _{\mathbf{x}t}f(\mathbf{x},t):=\frac{\partial}{\partial t}{\nabla}_\mathbf{x} f(\mathbf{x},t) \in \eucd ^{n}$, respectively.

\section{Differential Flatness}\label{sec:preliminaries}
Over the past several decades, differential flatness theory has been a main direction in the area of nonlinear control for motion planning, trajectory generation, and stabilization  \cite{levine2009analysis,van1998differential,murray1995differential,levine2011necessary}. 
Roughly speaking, flat systems are those systems that are equivalent to a controllable linear one, namely, a system made of chains of integrators of arbitrary length \cite{van1998differential}. 
More precisely, consider the system
\begin{align}\label{eq:system}
      \Dot{\mathbf{x}} &= f(\mathbf{x},\mathbf{u} ).
\end{align}
with $\mathbf{x} \in \eucd^n$ being its state and $\mathbf{u} \in \mathbb{R}^{m}$ its inputs. {The system \eqref{eq:system} is \emph{differentially flat} if there exist positive integers $r, k \in \mathbb{N}$ and functions $h$, $\varphi$, $\alpha$ such that a \textit{flat output} $\mathbf{y} \in \mathbb{R}^{m}$ is expressible from the state and input as \cite{bullo2019geometric}}
\begin{align}
    \mathbf{y} = h(\mathbf{x}, \mathbf{u}^{[r]} ),
\end{align}
{and, conversely, the state and input are recovered from $\mathbf{y}$ and finitely many of its derivatives as} 
\begin{equation}
\begin{aligned}
\mathbf{x} = \varphi(\mathbf{y}^{[k]}), \;\;\mathbf{u} = \alpha(\mathbf{y}^{[k]}). \label{eq:input-output relationship_flat_system}
\end{aligned}    
\end{equation}

For our purposes, the main advantage of using flat outputs in control system design is that doing so simplifies the process of generating input trajectories that satisfy certain constraints. Instead of designing complex controllers that directly manipulate the control inputs, one can design controllers that manipulate the flat outputs, and then use the algebraic relationships between the flat outputs and the control inputs to generate optimal state and control trajectories. Notably, many commonly used classes of systems in nonlinear control theory are differentially flat, for example fully actuated holonomic systems, mobile robots, and classical $n$-trailer systems. A complete characterization of differential flatness and a catalog of finite dimensional flat systems can be found in \cite{levine2009analysis,van1998differential,murray1995differential,levine2011necessary}.

Another important concept in nonlinear control theory is \textit{(dynamic) feedback linearization}, in which a nonlinear system is transformed into a linear system by a state diffeomorphism and a feedback transformation \cite{isidori2013nonlinear}. Although differential flatness and feedback linearization are related, they are not identical. In particular, while all state feedback-linearizable systems are differentially flat, a differentially flat system need not be state feedback linearizable, and need not be dynamic feedback linearizable everywhere in the state space. 
Differential flatness is a geometric property of a nonlinear system, independent of coordinate representation, and this can be exploited for nonlinear controller design \cite{martin2003flat}.

\section{Problem Statement}\label{sec:ProblemStatement} 
In this section, we formally state the problem of interest, together with some regularity assumptions needed in our derivations, and we introduce two motivating examples. 
Consider a differentially flat system described as in \eqref{eq:system}, along with the associated flat output
\begin{equation}\label{Eq:ODEFlat}
      \Dot{\mathbf{x}} = f(\mathbf{x},\mathbf{u} ), \quad \mathbf{y}= h(\mathbf{x},\mathbf{u}^{[r]}).
\end{equation}
We begin by formulating a time-varying optimization problem in the variable $\mathbf{y}$. To this end, let $t \geq  0$ be a continuous time index, and $f_0: \eucd ^m \times \eucd _+ \to \eucd$ be a time-varying objective function of the flat output $\mathbf{y}$. For $p \in \mathbb{Z}_{\geq 0}$ and $i \in [p]$, the maps $f_i : \eucd^m  \times \eucd_+ \to \eucd$ will encode $p$ time-varying inequality constraints on $\mathbf{y}$. We similarly consider $q \in \mathbb{Z}_{\geq 0}$ time-varying linear equality constraints of the form $ \mathbf{a}_j(t)^{\T}\mathbf{y} = b_j(t)$, which we stack into matrix form as $\mathbf{A}(t)\mathbf{y} = \mathbf{b}(t)$, where now $\mathbf{A}: \eucd_+ \to \eucd^{q \times m}$ and $\mathbf{b}: \eucd_+ \to \eucd^{q}$; we assume throughout that $q < m$.
This leads to the following constrained TV-O problem
\begin{equation}
    \begin{aligned}
     \mathbf{y}^*(t) := \arg\min_{\mathbf{y} \in \eucd^m} \;& f_0(\mathbf{y},t) \\
                    \textrm{s.t.} \quad & f_i(\mathbf{y},t) \leq 0, \quad i \in [p]  \\
                    &\mathbf{A}(t)\mathbf{y} = \mathbf{b}(t).\label{eq:Time_Varing_Optimization_Problem}
\end{aligned}
\end{equation}
We will assume that the minimizer $\mathbf{y}^*(t)$ of \eqref{eq:Time_Varing_Optimization_Problem} is unique for all $ t \geq 0$ (see Assumption \ref{ass:convexity}).
{Our goal is to generate a control input $\mathbf{u}(t)$ for \eqref{Eq:ODEFlat} such that, for some $\alpha>0$ independent of the initial condition, every initial condition admits a constant $C>0$ satisfying $\|\mathbf{y}(t)-\mathbf{y}^*(t)\|_2 \leq Ce^{-\alpha t}$ for all $t\geq0$, thus ensuring global attractivity of the output to the minimizer trajectory $\mathbf{y}^*(t)$ with an exponential convergence rate.}  

Various motion planning and control tasks can be encoded as instances of the TV-O problem \eqref{eq:Time_Varing_Optimization_Problem}. For example, if the positions of a controlled robot and a moving target are denoted by $\mathbf{y}(t)$ and $\mathbf{y}^d(t)$ respectively, then minimizing 
$f_0(\mathbf{y},t) = \| \mathbf{y}(t)- \mathbf{y}^d(t)\|_2$ encodes the task of tracking the target. 
{Similarly, planar formation constraints specified by a complex Laplacian \cite{lin2014distributed} can be written as $\mathbf{A}\mathbf{y}(t)=0$, where $\mathbf{y}(t)$ stacks the real planar coordinates and $\mathbf{A}$ is obtained by separating real and imaginary parts and removing redundant equations; see \cite{lin2014distributed} for more details.}   Furthermore, to ensure collision avoidance with respect to $O\in\mathbb{N}$ obstacles, a set of inequality constraints $\{\mathbf{a}_i(\mathbf{y})^{\T}\mathbf{z} - b_i(\mathbf{y}) \leq 0,\, i \in [O]\}$ can be enforced \cite{fazlyab2017prediction}; this will be discussed further in Section \ref{subsec: obstacal avoidance}.

{We first state smoothness, convexity, and constraint-regularity conditions commonly used in time-varying optimization \cite{simonetto2020time}. Each result specifies those required; additional assumptions are introduced in the relevant sections.}

\begin{ass}[Smoothness]\label{ass:smooth}
All functions $f_0,f_1,\ldots,f_p$ are infinitely differentiable in both $\mathbf{y}$ and $t$, and $\mathbf{A}$ and $\mathbf{b}$ are infinitely differentiable in $t$.
\end{ass}
{\noindent Infinite differentiability is assumed for expositional simplicity only: the constructions and proofs below use derivatives of the problem data up to order $k+1$, where $k$ is the relative degree of the flat system.}

\begin{ass}[Uniform strong convexity]\label{ass:convexity}
The objective function $f_0(\mathbf{y},t)$ is uniformly strongly convex in $\mathbf{y}$, i.e., $\nabla_{\mathbf{yy}} f_0(\mathbf{y},t) \succeq m_f\mathbf{I}_m$ for some $m_f > 0$, for all $\mathbf{y}$ and for all $t \geq 0$. The inequality constraint functions $f_i(\mathbf{y},t)$ are convex in $\mathbf{y}$ for all $t \geq 0$ and for all $i \in [p]$.
\end{ass}

\begin{ass}[Uniform Mangasarian-Fromowitz constraint qualification]\label{ass:MFCQ}
For {the} global minimizer $\mathbf{y}^*(t)$ of \eqref{eq:Time_Varing_Optimization_Problem}
\begin{enumerate}
    \item there exists a uniformly bounded $\Bar{\mathbf{d}}(t) \in \eucd^m$, i.e., $ \| \Bar{\mathbf{d}}(t) \|_2 \! \leq \! d$ for some constant $d > 0$, and a constant $\epsilon > 0$ such that for all $t \geq 0$
    \begin{align*}
        \nabla_{\mathbf{y}}f_i(\mathbf{y}^*(t),t)^{\T} \Bar{\mathbf{d}}(t) \leq -\epsilon, & \qquad i \in \mathbb{I}(\mathbf{y}^*(t)), \\
        \mathbf{a}_j(t)^{\T} \Bar{\mathbf{d}}(t) = 0, & \qquad j \in [q],
    \end{align*}
     where $\mathbb{I}(\mathbf{y}^*(t)) \!:=\!  \{i \in [p]\, |\,f_i(\mathbf{y}^*(t),t) \!= \!0 \} $ denotes the index set associated with active inequality constraints. 
    \item there exist constants $0 < \tau_{\rm min} \leq \tau_{\rm max} < +\infty$ such that $\sigma_{\rm min} (\mathbf{A}(t)) \geq \tau_{\rm min}$ and $\sigma_{\rm max} (\mathbf{A}(t)) \leq \tau_{\rm max}$ for all $t \geq 0$, i.e., the vectors $\{ \mathbf{a}_j\}$ for $j \in [q]$ are uniformly linearly independent and uniformly bounded.
    \end{enumerate}
\end{ass}

Since the time-varying convex optimization problem has smooth objective and constraints functions, Assumption \ref{ass:convexity} and Assumption \ref{ass:MFCQ} imply that for all $t \geq 0$,  the Karush-Kuhn-Tucker (KKT) conditions \cite{boyd2004convex} provide necessary and sufficient conditions for optimality.
{Intuitively, Assumption \ref{ass:MFCQ} asks for a uniformly good feasible direction at the optimizer: one along which every active inequality constraint strictly decreases while the equality constraints are preserved. Section \ref{sec:Numerical} discusses why this holds in both numerical examples discussed in this paper.} 
{The uniform margins preclude degenerating constraint qualifications as $t\to\infty$. Section~\ref{sec:Ineq Constrained TVO Framework} combines these conditions with additional regularity assumptions to prove uniform boundedness of the optimal multipliers, a property assumed in prior work \cite{fazlyab2017prediction}.}

{The remainder of this section develops a motivating example in two steps. We first illustrate our design procedure on the simplest flat system, an integrator, where matching the flat output to a time-varying gradient flow recovers the \emph{prediction--correction} methods of time-varying optimization \cite{fazlyab2017prediction,simonetto2016class}. We then extend the approach to a second-order nonholonomic system, the wheeled mobile robot, whose flat output must instead be matched to a second-order gradient algorithm. We close by summarizing the key features of the framework.} 

\subsection{Example \#1: Integrator}\label{subsec:Ex_1_Integrator}

{To fix ideas, consider first the simplest flat system:}
\begin{align}\label{eq:Integrator_System_Dynamic}
    \mathbf{\dot{x} = u}, \quad \mathbf{y = x}
    \;\;\Longleftrightarrow\;\;
    \mathbb{F}(\dot{\mathbf{y}},\mathbf{u}):= \mathbf{\dot{y} - u} = 0,
\end{align}
{{where $\mathbb{F}=0$ encodes the system input-output relation. We seek to track} the minimizer $\mathbf{y}^*(t)$ of the \emph{unconstrained} problem}
\begin{equation}\label{eq: Unconstrained TV optimization problem in example}
    \mathbf{y}^*(t) = \arg\min_{\mathbf{y} \in \eucd^{m}} \; f_0(\mathbf{y},t),
\end{equation}
{which, under Assumption \ref{ass:convexity}, is characterized by $\nabla_\mathbf{y} f_0(\mathbf{y}^*(t),t) = 0$. This motivates the \emph{target system}}
\begin{equation}
\dot \nabla_\mathbf{y} f_0(\mathbf{y},t) = -\mathbf{P}\nabla_\mathbf{y} f_0(\mathbf{y},t), \quad \mathbf{P}\succ 0,  \label{eq:Gradient-PositiveDefinite-System}
\end{equation}
{whose solutions satisfy $\nabla_{\mathbf{y}}f_0(\mathbf{y}(t),t) \to 0$ exponentially, and hence $\mathbf{y}(t) \to \mathbf{y}^*(t)$. {By the chain rule, the target system \eqref{eq:Gradient-PositiveDefinite-System} is equivalent to
\[
\begin{aligned}
\mathbb{G}(\mathbf{y},\dot{\mathbf{y}},t)
:= {}& \nabla_{\mathbf{yy}}f_0(\mathbf{y},t)\dot{\mathbf{y}}
     + \nabla_{\mathbf{y}t}f_0(\mathbf{y},t)\\
&+ \mathbf{P}\nabla_{\mathbf{y}}f_0(\mathbf{y},t)=0.
\end{aligned}
\]
Solving $\mathbb{G}=0$ for $\dot{\mathbf{y}}$ and $\mathbb{F}=0$ for $\mathbf{u}$ yields the feedback law}}
\begin{align}
   \mathbf{u}= -\nabla_{\mathbf{yy}}^{-1}f_0(\mathbf{y},t)[\mathbf{P}\nabla_\mathbf{y} f_0(\mathbf{y},t)
    +\nabla_{\mathbf{y}t}f_0(\mathbf{y},t)], \label{eq:integrator-control}
\end{align}
{which is well defined by Assumption \ref{ass:convexity}.}

{The two terms of \eqref{eq:integrator-control} recover the \emph{prediction} and \emph{correction} steps of prediction--correction methods in time-varying optimization \cite{fazlyab2017prediction,simonetto2016class}, and the resulting closed loop is the linear gradient system $\dot{\mathbf{w}} = -\mathbf{Pw}$ with $\mathbf{w} = \nabla_\mathbf{y} f_0(\mathbf{y},t)$ (cf.\ Fig.~\ref{fig:framework}). The integrator is special, however, in that $\dot{\mathbf{y}}$ is directly assignable.}

\subsection{Example \#2: Wheeled Mobile Robot}\label{subsec:Ex_2_WMR}

{We now turn to a system where $\dot{\mathbf{y}}$ is not directly assignable, in particular,}  the wheeled mobile robot (WMR)~\cite{martin2003flat}:
\begin{equation}\label{eq:Unicycle_System_Dynamic}
\begin{aligned}
\Dot{x}_1 &=\cos(x_3)u_1,
&\Dot{x}_2&=\sin(x_3)u_1,\\ 
\Dot{x}_3 &=u_2, 
& \mathbf{y} &= (x_1,x_2).
\end{aligned}
\end{equation}
The states $(x_1,x_2)\in\mathbb{R}^2$ represent the position, and $x_3$ is the angular position of the WMR. The control inputs $(u_1, u_2)$ are the positional and angular velocities, respectively, and the position vector $\mathbf{y}$ is the flat output. Consider again the unconstrained time-varying optimization problem \eqref{eq: Unconstrained TV optimization problem in example}, where our goal is now to regulate the output vector $\mathbf{y}$ of the WMR to asymptotically track the time-varying minimizer. 

For the WMR \eqref{eq:Unicycle_System_Dynamic}, the algebraic relationship \eqref{eq:input-output relationship_flat_system} between the flat output and the input is
\begin{equation}\label{Eq:WMR-u}
    \mathbf{u} := \alpha(\mathbf{y}^{[k]})= \begin{bmatrix}
        \sqrt{\dot{y}_1^2 + \dot{y}_2^2 } \\
        (\dot{y}_1\ddot{y}_2 - \ddot{y}_1\dot{y}_2) / (\dot{y}_1^2 + \dot{y}_2^2)
    \end{bmatrix}.
\end{equation}
which we express implicitly as
\begin{align}
\mathbb{F}(\mathbf{y},\dot{\mathbf{y}},\ddot{\mathbf{y}},\mathbf{u}):=
\begin{bmatrix}
    u_1 - \sqrt{\dot{y}_1^2 + \dot{y}_2^2 } \nonumber \\
    u_2 - (\dot{y}_1\ddot{y}_2 - \ddot{y}_1\dot{y}_2) / (\dot{y}_1^2 + \dot{y}_2^2)
\end{bmatrix}= \begin{bmatrix}0 \\ 0 \end{bmatrix}.
\end{align}  
{Note that $u_2$ depends on $\ddot{\mathbf{y}}$: the WMR has relative degree $k=2$, so the controller cannot assign $\dot{\mathbf{y}}$ directly, as the first-order design above would require, and must instead shape the dynamics of the gradient itself, driving both $\nabla_{\mathbf{y}}f_0$ and its time derivative $\dot{ \nabla}_\mathbf{y} f_0$ to zero. This is the natural setting for mechanical systems, whose force- or torque-actuated dynamics have relative degree two in position outputs.}

Accordingly, we generalize the first-order target system \eqref{eq:Gradient-PositiveDefinite-System} to the second-order target system
\begin{align}
    \begin{bmatrix}
    \dot{ \nabla}_\mathbf{y} f_0(\mathbf{y},t) \\ \ddot{ \nabla}_\mathbf{y} f_0(\mathbf{y},t)
    \end{bmatrix} = \begin{bmatrix}
     0 & \mathbf{I}_m \\ -k_{\rm p} \mathbf{I}_m  &-k_{\rm d} \mathbf{I}_m
      \end{bmatrix} \begin{bmatrix}
       \nabla_\mathbf{y} f_0(\mathbf{y},t) \\ \dot{\nabla}_\mathbf{y} f_0(\mathbf{y},t)
\end{bmatrix},\label{eq:Second_order_Optimality_Error} 
\end{align}
where $k_{\rm p}, k_{\rm d} > 0$. Equation \eqref{eq:Second_order_Optimality_Error} defines an exponentially stable linear system with state $\mathbf{z}:=(\nabla_\mathbf{y} f_0(\mathbf{y},t),\dot{\nabla}_\mathbf{y} f_0(\mathbf{y},t))$. To determine the required target evolution of $\mathbf{y}$, we differentiate the gradient term $\nabla_\mathbf{y} f_0(\mathbf{y},t)$ with respect to time twice, yielding
\begin{align*}     \ddot{\nabla}_{\mathbf{y}}f_0(\mathbf{y},t) = &\nabla_{\mathbf{yy}}f_0(\mathbf{y},t)\mathbf{\ddot{y}}+ \dot{\nabla}_{\mathbf{yy}}f_0(\mathbf{y},t)\mathbf{\dot{y}}+\dot{ \nabla}_{\mathbf{y}t}f_0(\mathbf{y},t). 
\end{align*} 
Equating this with the second row of \eqref{eq:Second_order_Optimality_Error}, we obtain the implicit equation
\begin{equation}
\begin{aligned}
\mathbb{G}(\mathbf{y},\dot{\mathbf{y}},\ddot{\mathbf{y}},t) := \nabla_{\mathbf{yy}}f_0(\mathbf{y},t)\mathbf{\ddot{y}}+ \dot{\nabla}_{\mathbf{yy}}f_0(\mathbf{y},t)\mathbf{\dot{y}} \\ + \dot{ \nabla}_{\mathbf{y}t}f_0(\mathbf{y},t) + k_{\rm p}\nabla_\mathbf{y} f_0(\mathbf{y},t) +k_{\rm d} \dot{\nabla}_\mathbf{y} f_0(\mathbf{y},t) = 0, \label{eqn:WMR_ODE_Time-varying_Optimization}
\end{aligned}    
\end{equation}
which describes the solution trajectory of the time-varying optimization problem.  

We now have the two implicit equations $ \mathbb{F}(\mathbf{y},\dot{\mathbf{y}},\ddot{\mathbf{y}},\mathbf{u}) = 0$ and $\mathbb{G}(\mathbf{y},\dot{\mathbf{y}},\ddot{\mathbf{y}},t)=0$, and seek a simultaneous solution of the form $(\ddot{\mathbf{y}}, \mathbf{u})= S(\mathbf{y},\dot{\mathbf{y}} ,t)$. In this case, we may explicitly solve \eqref{eqn:WMR_ODE_Time-varying_Optimization} for $\ddot{\mathbf{y}}$ and substitute into \eqref{Eq:WMR-u}, yielding the solution
\begin{gather*}
\begin{aligned}
\ddot{\mathbf{y}} &= g(\mathbf{y}, \dot{\mathbf{y}}) := -\nabla_{\mathbf{yy}}^{-1}f_0(\mathbf{y},t)\big[\dot{\nabla}_{\mathbf{yy}}f_0(\mathbf{y},t)\mathbf{\dot{y}}+\dot{ \nabla}_{\mathbf{y}t}f_0(\mathbf{y},t) \\
   & \qquad\qquad +k_{\rm p}\nabla_\mathbf{y} f_0(\mathbf{y},t)+k_{\rm d}\dot{\nabla}_\mathbf{y} f_0(\mathbf{y},t)\big]
\end{aligned}\\
u_1 = \|\mathbf{\dot{ y}}\|_2 ,\qquad
u_2 = \frac{1}{ \|\mathbf{\dot{ y}}\|_2^2}\,g(\mathbf{y}, \dot{\mathbf{y}})^{\T} {\begin{bmatrix}
0 &-1\\
1 &0
\end{bmatrix}}\dot{\mathbf{y}}.
\end{gather*}

This feedback law again generalizes feedback linearization: it transforms the WMR \eqref{eq:Unicycle_System_Dynamic} into the optimization algorithm
\[
    \mathbf{\dot{z} = H z}, \qquad \mathbf{z} = ( \nabla_\mathbf{y} f_0(\mathbf{y},t),\dot{\nabla}_\mathbf{y} f_0(\mathbf{y},t))^{\T},
\]
where $\mathbf{H}$ is the Hurwitz matrix from \eqref{eq:Second_order_Optimality_Error}. This optimization algorithm seeks to find the optimal solution of the unconstrained version of the time-varying optimization problem \eqref{eq:Time_Varing_Optimization_Problem}.

\begin{figure}[hbt!]
\centerline{\includegraphics[width=0.85\columnwidth]{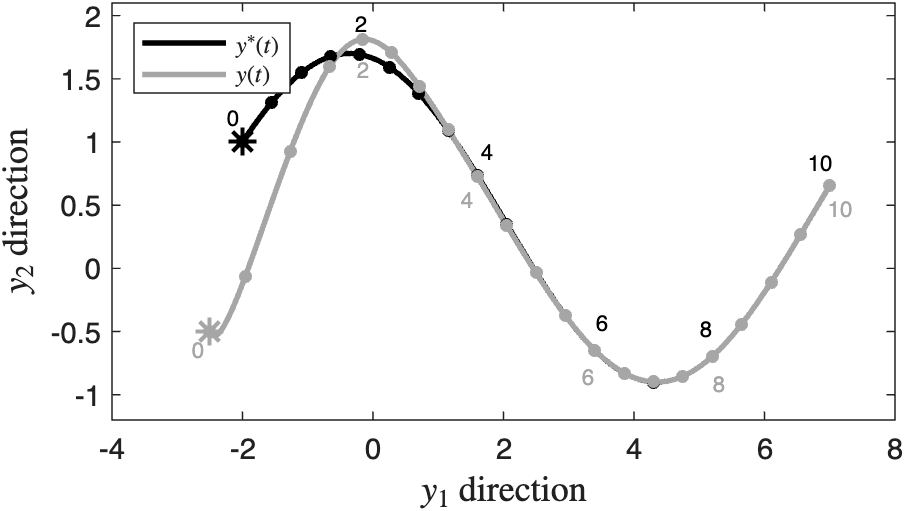}}
   \caption{Plot of a robot tracking an object, where $\mathbf{y}^*(t)$ \eqref{eq:Time_Varing_Optimization_Problem} is the minimizer and $\mathbf{y}(t)$ represents the real trajectory of the robot, where the dots represent their locations at different $t$. Due to mismatching initial conditions (highlighted using asterisk), we design a control law that converges asymptotically to $\mathbf{y}^*(t)$.}
\label{fig:2}
\end{figure}

Figure \ref{fig:2} illustrates the behavior of the proposed solution. Notably, even when $\mathbf{y}^*(t)$ is not a feasible trajectory due to, e.g., a mismatch on the initial conditions, the control law forces the trajectory to asymptotically converge to $\mathbf{y}^*(t)$. 

{\begin{rem}[Scope of the constraints]\label{rem:constraint scope}
The constraints in our formulation act on the flat output $\mathbf{y}$ and, via \eqref{eq:input-output relationship_flat_system}, on quantities expressible in $\mathbf{y}$ and its derivatives. General state and input constraints can in principle be rewritten in this form, but typically yield nonconvex functions outside our assumptions. Their systematic treatment is left as future work. Input limits are examined empirically in Section \ref{sec:Numerical}, enabled by the relaxed barrier of Section \ref{sec:Ineq Constrained TVO Framework}.
\end{rem}}

\subsection{Key Features}\label{subsec:Key Features}
{Both examples combine the system input-output relation $\mathbb{F}(\mathbf{y}^{[k]},\mathbf{u})=0$ with the optimization dynamics $\mathbb{G}(\mathbf{y}^{[k]},t)=0$. Given $\mathbf{y}^{[k-1]}$ and $t$, we seek the highest output derivative and the input in the form
\[
(\mathbf{y}^{(k)},\mathbf{u})=S(\mathbf{y}^{[k-1]},t).
\]}

For a flat system, \eqref{eq:input-output relationship_flat_system} gives the system dynamics term directly, $\mathbb{F} (\mathbf{y}^{[k]},{\mathbf{u}}):= \mathbf{u} -\alpha(\mathbf{y}^{[k]} ) = 0$, so the subsequent sections focus on the optimization dynamics term: in both examples, $\mathbb{G}(\mathbf{y}^{[k]} ,t) = 0$ was obtained from an exponentially stable linear system whose state comprises the gradient $\nabla_\mathbf{y} f_0(\mathbf{y},t)$ and its time derivatives, cf.\ \eqref{eqn:WMR_ODE_Time-varying_Optimization}.

{The rest of the paper generalizes this construction: given an arbitrary flat system \eqref{eq:system} and problem \eqref{eq:Time_Varing_Optimization_Problem}, we design $\mathbb{G}(\mathbf{y}^{[k]},t) = 0$ so that every solution $\mathbf{y}(t)$ globally asymptotically converges to the minimizer.}  
A key to the success of this effort is the design of general target systems of the form
\begin{equation}
    \label{eq:transformation algorithm TVO}
    \mathbf{\dot{w} = H w},\quad
    \mathbf{w} = (\nabla_\mathbf{z} L(\mathbf{z},t),...,\nabla^{k-1}_\mathbf{z} L(\mathbf{z},t))^{\T},
\end{equation}
where $L$, $\mathbf{z}$, and $\mathbf{H}$, will be properly chosen to guarantee the asymptotic convergence of $\mathbf{y}(t)$ to the optimal solution of the general \emph{constrained} time-varying optimization problem  \eqref{eq:Time_Varing_Optimization_Problem}.

\section{Unconstrained Time-varying Optimization Framework}\label{sec:Unconstrained_TV_optimization_Framework}
In this section, we first consider the case where our goal is to regulate a differentially flat system \eqref{eq:system} to the minimizer $\mathbf{y}^*(t)$ of 
the unconstrained time-varying optimization problem 
\begin{align}\label{eq: Unconstrained Time_Varing_Optimization_Problem}
     \mathbf{y}^*(t) := &\arg\min_{\mathbf{y} \in \eucd^m} \; f_0(\mathbf{y},t). 
\end{align} 

Recall that for a differentially flat system \eqref{eq:system}, the inputs are determined by the flat outputs and a finite number of their derivatives according to \eqref{eq:input-output relationship_flat_system}. Put differently, the implicit function $  \mathbb{F} (\mathbf{y}^{[k]},{\mathbf{u}}):= \mathbf{u} -\alpha(\mathbf{y}^{[k]} ) =  0$ encodes the system dynamics, which is a function of up to $k$-th order derivatives of the flat output $ \mathbf{y}$.
The time-varying optimization problem is assumed to be uniformly convex (Assumption \ref{ass:convexity}) and the objective smooth in both $\mathbf{y}$ and $t$. 
Building on the previously considered target systems \eqref{eq:Gradient-PositiveDefinite-System} and \eqref{eq:Second_order_Optimality_Error}, we now consider the $k$-th order target system
\begin{align}
    \begin{bmatrix}
       \dot{\nabla}_\mathbf{y} f_0(\mathbf{y},t) \\ \vdots \\ \nabla^{(k)}_\mathbf{y} f_0(\mathbf{y},t)
    \end{bmatrix} = \mathbf{H}
    \begin{bmatrix}
       \nabla_\mathbf{y} f_0(\mathbf{y},t) \\ \vdots \\ \nabla^{(k-1)}_\mathbf{y} f_0(\mathbf{y},t)
    \end{bmatrix}   ,\label{eq:K-th_order_Unconstrained_Optimality_Error}
\end{align} where
\begin{align*}
    \mathbf{H} = \hat{\mathbf{H}} \otimes \mathbf{I}_m, \;\;  \hat{\mathbf{H}} :=  \begin{bmatrix}
    0 &1 &0 &\dots &0 \\
    0 &0 &1 &\dots &0 \\
    \vdots & \vdots &  &\ddots & \vdots\\
    -a_0 & -a_1& -a_2 &\dots &-a_{k-1}
    \end{bmatrix} \label{eq:Hurwitz_Matrix_Big_H}
\end{align*}
is Hurwitz. 
Equation \eqref{eq:K-th_order_Unconstrained_Optimality_Error} is the natural generalization of \eqref{eq:Gradient-PositiveDefinite-System} and \eqref{eq:Second_order_Optimality_Error}. The need to increase the order of the target system as the order of the flat system increases is evidenced by the following lemma.
\begin{lem}[$k$-th order time derivative of $\nabla_\mathbf{y} f_0(\mathbf{y},t)$] \label{lem:Diffrentiating_Gradient_K_Times}
The $k$th total time derivative of $\nabla_\mathbf{y} f_0(\mathbf{y},t)$ can be expressed as
\begin{equation}
\begin{aligned}
    \nabla_\mathbf{y}^{(k)} f_0(\mathbf{y},t) &= \sum_{m = 0}^{k-1} {\textstyle \binom{k-1}{m}} \nabla_{\mathbf{yy}}^{(m)}f_0(\mathbf{y},t)\mathbf{y}^{(k-m)}\\
    &\quad + \nabla_{\mathbf{y}t}^{(k-1)}f_0(\mathbf{y},t),
\end{aligned} \label{eq:Differentiate_gradient_K_Times}
\end{equation}
where $\binom{k-1}{m}$ denotes the binomial coefficient.
\end{lem}

 \begin{proof}
 See Appendix~\ref{Appendix:Diffrentiating_Gradient_K_Times}.\end{proof}

\smallskip

{Lemma \ref{lem:Diffrentiating_Gradient_K_Times} shows that $k$ is the smallest number of differentiations of $\nabla_{\mathbf{y}}f_0(\mathbf{y},t)$ needed for $\mathbf{y}^{(k)}$ to appear. The derivative $\mathbf{y}^{(k)}$ is the highest output derivative required by \eqref{eq:input-output relationship_flat_system} to compute the control. Thus, combining \eqref{eq:K-th_order_Unconstrained_Optimality_Error} and \eqref{eq:Differentiate_gradient_K_Times}, for a general flat system and unconstrained time-varying optimization problem, we obtain the implicit equation}  
\begin{equation}
    \begin{aligned}
\mathbb{G}_{\mathrm{unc}}(\mathbf{y}^{[k]}, t) := \sum_{m=0}^{k-1}{\textstyle\binom{k-1}{m}} \nabla_{\mathbf{yy}}^{(m)}f_0(\mathbf{y},t)\mathbf{y}^{(k-m)}  \\+  \nabla_{\mathbf{y}t}^{(k-1)}f_0(\mathbf{y},t) 
+ \sum_{i = 0} ^{k-1} a_i\nabla_{\mathbf{y}}^{(i)}f_0(\mathbf{y},t) = 0.\label{eq:design_requirement_q_order_unconstr}
\end{aligned}
\end{equation}
which specifies the desired \emph{optimization dynamics}. The next result states that if certain regularity conditions are met, the solutions $\mathbf{y}(t)$ of the optimization dynamics described in equation \eqref{eq:design_requirement_q_order_unconstr} will asymptotically converge to the minimizer $\mathbf{y}^*(t)$ of equation \eqref{eq: Unconstrained Time_Varing_Optimization_Problem}.

\begin{thm}[Convergence of optimization dynamics \eqref{eq:design_requirement_q_order_unconstr}] \label{Thm: Unconstrained} Let Assumptions \ref{ass:smooth} and \ref{ass:convexity} hold. Then for any initial condition, the trajectory $t\mapsto \mathbf{y}(t)$ of system  $\mathbb{G}_{\mathrm{unc}}(\mathbf{y}^{[k]},t) = 0$ defined in   \eqref{eq:design_requirement_q_order_unconstr}  globally asymptotically converges to the optimal solution $\mathbf{y}^*(t)$ of \eqref{eq: Unconstrained Time_Varing_Optimization_Problem}.
Moreover, the estimates
\begin{align*}
    & \| \mathbf{y}(t)-\mathbf{y}^*(t)\|_2 \leq Ce^{-\alpha t} ,\\
    & f_0(\mathbf{y}(t),t) - f_0(\mathbf{y}^*(t),t) \leq m_fC^2e^{-2\alpha t}
\end{align*}
hold, for all $t\ge0$, where 
\[
C = \left(\tfrac{c^2}{m_f^2} \sum_{j=0}^{k-1}\nolimits \|{\nabla_\mathbf{y}^{(j)} f_0(\mathbf{y}(0),0)}\|_2^2)\right)^{\frac{1}{2}} < \infty,
\]
for some constant $c > 0$ , and where $-\alpha := \max_{\lambda\in\spec(\mathbf{H})} \Re[\lambda]+ \epsilon_H $, for some  $ \epsilon_H >0$ sufficiently small.
\end{thm}
 \begin{proof}
 See Appendix~\ref{Appendix: thm Unconstrained}.\end{proof}

The above theorem states that the solution trajectory of $ \mathbb{G}_{\mathrm{unc}}(\mathbf{y}^{[k]}, t) =0$ from \eqref{eq:design_requirement_q_order_unconstr} converges asymptotically to the minimizer $\mathbf{y}^*(t)$ of \eqref{eq: Unconstrained Time_Varing_Optimization_Problem}. It remains to show that one can indeed simultaneously resolve the system of equations $ \mathbb{F} (\mathbf{y}^{[k]},{\mathbf{u}})=0$ and $\mathbb{G}_{\mathrm{unc}}(\mathbf{y}^{[k]}, t) =0 $
 for a solution $ (\mathbf{y}^{(k)},  \mathbf{u} ) = S(\mathbf{y}^{[k-1]},t)$, {yielding the control law achieving this.}
By uniform strong convexity (see Assumption \ref{ass:convexity}), the Hessian matrix $\nabla_{\mathbf{yy}} f_0(\mathbf{y},t) $ is uniformly positive definite. This approach allows us to solve for $ (\mathbf{y}^{(k)},  \mathbf{u} )$ by first determining $\mathbf{y}^{(k)}$ using \eqref{eq:design_requirement_q_order_unconstr}, and then solving for $\mathbf{u}$ using \eqref{eq:input-output relationship_flat_system}. The following theorem summarizes these findings. 

\begin{thm}[TVO-based control for system \eqref{eq:system}]\label{Thm: Unconstrained_Control}
Let Assumptions \ref{ass:smooth} and \ref{ass:convexity} hold and consider the differentially flat system  \eqref{eq:system} with the feedback controller
\[
\mathbf{u} := \alpha(\mathbf{y},\dots, \mathbf{y}^{(k-1)}, g_{\mathrm{unc}}(\mathbf{y}^{[k-1]} ))
\]
where
\[
\begin{aligned}
   &g_{\mathrm{unc}}(\mathbf{y}^{[k-1]} )  
   =-\nabla_{\mathbf{yy}}^{-1}f_0(\mathbf{y},t)\left[ \sum_{i = 0} ^{k-1} a_i\nabla_{\mathbf{y}}^{(i)}f_0(\mathbf{y},t) \right.\\
  & \quad +  \nabla_{\mathbf{y}t}^{(k-1)}f_0(\mathbf{y},t) 
  + \left. \sum_{m=1}^{k-1}{\textstyle{\binom{k-1}{m}}  }\nabla_{\mathbf{yy}}^{(m)}f_0(\mathbf{y},t)\mathbf{y}^{(k-m)}  \right] 
\end{aligned}
\]
Then, for any initial condition, the flat output of 
\eqref{eq:system} globally asymptotically converges to the optimal solution $\mathbf{y}^*(t)$ of \eqref{eq: Unconstrained Time_Varing_Optimization_Problem}.
\end{thm}
{\begin{proof} The proof is immediate from the preceding constructions: Lemma \ref{lem:Diffrentiating_Gradient_K_Times} and Assumption \ref{ass:convexity} show that \eqref{eq:design_requirement_q_order_unconstr} uniquely defines $\mathbf{y}^{(k)} = g_{\mathrm{unc}}(\mathbf{y}^{[k-1]})$, which the input $\mathbf{u} = \alpha(\mathbf{y},\dots,\mathbf{y}^{(k-1)},g_{\mathrm{unc}})$ realizes exactly by flatness; convergence then follows from Theorem \ref{Thm: Unconstrained}. \end{proof}}

The nonlinear feedback in Theorem \ref{Thm: Unconstrained_Control} effectively transforms the differentially flat system into the optimization algorithm \eqref{eq:transformation algorithm TVO} which then converges to the optimal solution of the time-varying optimization problem.

{\begin{rem}[The closed loop as the solver]\label{rem:solver}

Contrast the design of Theorem \ref{Thm: Unconstrained_Control}  with first solving \eqref{eq: Unconstrained Time_Varing_Optimization_Problem} for $\mathbf{y}^*(t)$ and tracking the result. Since the problem is time-varying, that alternative must either precompute $\mathbf{y}^*(\cdot)$ offline, which requires the problem data in advance, or re-solve the time-frozen problem at each sampling instant and numerically differentiate the computed samples to obtain the derivatives that flatness-based tracking requires. The controller of Theorem \ref{Thm: Unconstrained_Control} does neither: the closed loop itself is the optimization algorithm, evaluating analytic derivatives of the problem data at the current time, so only causal access to the data is needed. Moreover, no pre-stabilization is required (flat systems have no zero dynamics), and, under Assumptions \ref{ass:smooth} and \ref{ass:convexity}, for stationary problems the closed loop renders $(\mathbf{y}^*, 0, \dots, 0)$ a globally asymptotically stable equilibrium (the time-invariant case of Theorem \ref{Thm: Unconstrained}).
\end{rem}}

\section{Equality Constrained Time-varying Optimization Framework}\label{sec:Equality Constrained TVO Framework}
In this section, we consider the equality-constrained version of the time-varying optimization problem \eqref{eq:Time_Varing_Optimization_Problem}, written as
\begin{equation}
    \begin{aligned}\label{eq:Eq constrained TVO}
        \mathbf{y}^*(t) := \arg\min_{\mathbf{y} \in \eucd^m} \; & f_0(\mathbf{y},t) \\
                    \textrm{s.t.}\quad   &\mathbf{A}(t)\mathbf{y} = \mathbf{b}(t).
    \end{aligned}
\end{equation}
Define the Lagrangian ${L}: \eucd^m \times \eucd^q \times \eucd_+ \to \eucd$ associated with the problem as 
\begin{equation}
    {L}(\mathbf{y},\boldsymbol{\nu},t) = f_0(\mathbf{y},t) + \boldsymbol{\nu}^{\T} (\mathbf{A}(t)\mathbf{y} - 
    \mathbf{b}(t)),
\end{equation}
and where $\boldsymbol{\nu}_i$ as the Lagrange multiplier associated with the $i$th equality constraint $\mathbf{a}_i(t)^{\T} \mathbf{y} = b_i(t)$. 
{The construction of Section \ref{sec:Unconstrained_TV_optimization_Framework} applies verbatim with $f_0$ replaced by $L$.}
The Assumption \ref{ass:MFCQ} on $\mathbf{A}(t)$ implies that there are fewer equality constraints than primal variables, and that the equality constraints are uniformly independent. Additionally, the time-varying optimization problem is uniformly convex (see Assumption \ref{ass:convexity}), and the KKT conditions are necessary and sufficient for the points to be primal and dual optimal. The optimal trajectory $\mathbf{z}^*(t) = \mathrm{col} (\mathbf{y}^*(t), \boldsymbol{\nu}^*(t)) \in \eucd^{m+q} $ is therefore characterized by the KKT conditions $\nabla_{\mathbf{z}} {L}(\mathbf{z}^*(t) ,t)= 0$, or more explicitly
\[
\begin{aligned}
    0 &= \nabla_{\mathbf{y}} {L}(\mathbf{y}^*(t) ,\boldsymbol{\nu}^*(t) ,t) = \nabla_{\mathbf{y}} f_0(\mathbf{y}^*(t),t) + \mathbf{A}(t)^{\T} \boldsymbol{\nu}^*(t)\\
    0 &= \nabla_{\boldsymbol{\nu}} {L}(\mathbf{y}^*(t) ,\boldsymbol{\nu}^*(t) ,t) = \mathbf{A}(t)\mathbf{y}^*(t)  - \mathbf{b}(t).
 \end{aligned}
 \]
Building upon \eqref{eq:K-th_order_Unconstrained_Optimality_Error}, the idea is to ensure that $\mathbf{z}(t)$ converges to the optimal primal-dual trajectory $\mathbf{z}^*(t)$ by designing a linear target system $\dot{\mathbf{w}} = \mathbf{H}\mathbf{w}$ with state $\mathbf{w} = \mathrm{col}(\nabla_{\mathbf{z}} {L}(\mathbf{z},t),\dots,\nabla^{(k-1)}_{\mathbf{z}} {L}(\mathbf{z},t))$, i.e., 
\begin{align}
    \begin{bmatrix}
       \dot{\nabla}_{\mathbf{z}} {L}(\mathbf{z},t) \\ \vdots \\ \nabla^{(k)}_\mathbf{{z}} {L}(\mathbf{z},t)
    \end{bmatrix} = \mathbf{H}
    \begin{bmatrix}
       \nabla_{\mathbf{z}} {L}(\mathbf{z},t) \\ \vdots \\ \nabla^{(k-1)}_{\mathbf{z}} {L}(\mathbf{z},t)
    \end{bmatrix}   \label{eq:K-th_order_equality_constraint_Optimality_Error},
\end{align} where $ \mathbf{H} = \hat{\mathbf{H}} \otimes \mathbf{I}_{m+q} $ is Hurwitz. As a result of Lemma \ref{lem:Diffrentiating_Gradient_K_Times}, differentiating the gradient $\nabla_\mathbf{z} {L}(\mathbf{z},t)$ with respect to time $k-$times yields
\begin{equation}
    \nabla_\mathbf{z}^{(k)}\!{L}\!(\mathbf{z},\!t\!) \!=\! \sum_{m = 0}^{k-1}\!{\textstyle \binom{k-1}{m}}\!\nabla_{\mathbf{zz}}^{(m)}\!{L}\!(\mathbf{z},\!t\!)\mathbf{z}^{(k-m)}
   \! + \!\nabla_{\mathbf{z}t}^{(k-1)}\!{L}\!(\mathbf{z},\!t\!).
\label{eq:Differentiate_lagrange_gradient_K_Times}
\end{equation}
In particular $\nabla_{\mathbf{zz}}{L}(\mathbf{z},t)$ is the KKT matrix \cite{boyd2004convex}
\begin{align*} 
    \nabla_{\mathbf{zz}}{L}(\mathbf{z},t) = \begin{bmatrix} \nabla_{\mathbf{yy}} f_0(\mathbf{y},t) & \mathbf{A}(t)^{\T} \\
    \mathbf{A}(t) & \mathbf{0}_{q \times q}
    \end{bmatrix},
\end{align*}
which is nonsingular because $\nabla_{\mathbf{yy}} f_0(\mathbf{y},t) \succeq m_f I_{m}$ and $\mathrm{rank} (\mathbf{A}(t)) = q $ for all $t \geq 0$ by Assumptions \ref{ass:convexity} and  \ref{ass:MFCQ}. This also means the optimal primal-dual pair $(\mathbf{y}^*(t), \boldsymbol{\nu}^*(t))$ is unique at each $t \geq 0$. 
However, our convergence analysis will require a uniform lower bound on the eigenvalues of $\nabla_{\mathbf{zz}}{L}(\mathbf{z},t)$, which in turn ensures the uniform bound $\|\nabla^{-1}_{\mathbf{zz}}{L}(\mathbf{z},t) \|_2 \leq K^{-1}$. Such bounds appear in the analysis of algorithms such as Newton and interior point methods, and serve a role similar to that of strong convexity (Assumption \ref{ass:convexity}) for unconstrained settings \cite{boyd2004convex,wright2001convergence,armand2013uniform,greif2014bounds}. {For the equality-constrained formulation, we additionally impose the following Lipschitz-gradient condition to bound the inverse KKT matrix.}

\begin{ass}[Uniform Lipschitz Continuous Gradient] \label{ass:lipschitz_gradient}
The objective function $f_0(\mathbf{y},t)$ has uniform Lipschitz Continuous Gradient, i.e., \begin{align}
        \|\nabla_{\mathbf{y}} f_0(\mathbf{y}_1,t) - \nabla_{\mathbf{y}} f_0(\mathbf{y}_2,t)\| _2\leq {L_f} \|\mathbf{y}_1 - \mathbf{y}_2 \|_2
\end{align}
for some ${L_f} > 0$, for all $\mathbf{y}_1,\mathbf{y}_2 \in  \eucd^m$ and for all $t \geq 0$.

\end{ass}

We will also use the following lemma to characterize the uniform boundedness of the eigenvalues of the KKT matrix. The lemma is stated for fixed (time-invariant) matrices, but will be later generalized for time-varying settings.

\begin{lem}[Eigenvalues of KKT matrix]\label{lem:Eigenvalues of KKT matrix}\cite[Lemma 2.1]{rusten1992preconditioned} 
    Suppose a matrix $\mathbf{A} $ takes the form 
    \begin{align*} 
    \mathbf{A} :=\begin{bmatrix}  \mathbf{M} & \mathbf{B}^{\T} \\
    \mathbf{B} & \mathbf{0}
    \end{bmatrix} \in {\eucd^{(m+q) \times (m+q)}},
\end{align*}
where $ \mathbf{M} \in \eucd^{m \times m}$ is symmetric and positive definite, $ \mathbf{B} \in  \eucd^{q \times m}$ with $q< m$ is full rank. Let $\mu_1\geq \mu_2\geq \cdots \geq \mu_m > 0$ be the eigenvalues of  $\mathbf{M}$, let $\sigma_1 \geq \sigma_2\geq \cdots \geq \sigma_q > 0$ be the singular values of $\mathbf{B}$, and denote by ${\spec(\mathbf{A})}$ the spectrum of the  matrix. Then ${\spec(\mathbf{A})}\subset I = I^-\cup I^+$, where $I^- = [\frac{1}{2}(\mu_m - \sqrt{\mu_m^2 + 4\sigma_1^2}), \frac{1}{2}(\mu_1 - \sqrt{\mu_1^2 + 4\sigma_q^2})]$ and $ I^+ = [ \mu_m, \frac{1}{2}(\mu_1 + \sqrt{\mu_1^2 + 4\sigma_1^2})]$.
\end{lem}

As mentioned before, the above Lemma bounds the spectrum of any fixed KKT matrix to be within two intervals, a negative $I^{-}$ and positive interval $I^{+}$. Notably, when $\mu_m>0$  and $\sigma_q>0$, the entire spectrum of the KKT matrix is bounded away from zero. A simple generalization of this argument for time-varying matrices establishes the uniform boundedness (for all times) of the inverse KKT matrix $\nabla_{\mathbf{zz}}L(\mathbf{z},t)$.
\begin{lem}[Uniform boundeness of inverse KKT matrix]\label{cor: uniform boundedness of inverse KKT}
    Under Assumptions~\ref{ass:smooth}--\ref{ass:lipschitz_gradient}, for all $t \geq 0$ and $\mathbf{z}$,
{\[
\|\nabla^{-1}_{\mathbf{zz}}L(\mathbf{z},t)\|_2\leq \tfrac{1}{K},\;
K:=\min\bigl\{m_f,\tfrac{\sqrt{L_f^2+4\tau_{\rm min}^{\,2}}-L_f}{2}\bigr\}>0.
\]}

\begin{proof}
 See Appendix~\ref{proof: cor: uniform boundedness of inverse KKT}.\end{proof}
\end{lem}

We are now ready to extend our framework to the equality-constrained problem \eqref{eq:Eq constrained TVO}. Combining \eqref{eq:K-th_order_equality_constraint_Optimality_Error} and \eqref{eq:Differentiate_lagrange_gradient_K_Times}, we use the following implicit function to define the \textit{optimization dynamics}, when time-varying equality constraints are included:
\begin{align}\label{eq:equality z_kth_lagrange_derivative_solution}
\mathbb{G}_{\mathrm{eq}}(\mathbf{z}^{[k]}, t) := \sum_{m=0}^{k-1}\binom{k-1}{m} \nabla_{\mathbf{zz}}^{(m)}{L}(\mathbf{z},t)\mathbf{z}^{(k-m)} \nonumber \\ +  \nabla_{\mathbf{z}t}^{(k-1)}{L}(\mathbf{z},t)  + \sum_{i = 0} ^{k-1} a_i\nabla_{\mathbf{z}}^{(i)}{L}(\mathbf{z},t) = 0. 
\end{align}
The following theorem states that if certain regularity conditions are met, the output $\mathbf{z} = (\mathbf{y},\mathbf{\nu})$, which follows the optimization dynamics described in equation \eqref{eq:equality z_kth_lagrange_derivative_solution}, will globally asymptotically converge to the optimal primal-dual solution $(\mathbf{y}^*(t),\boldsymbol{\nu}^*(t))$ of \eqref{eq:Eq constrained TVO}.

\begin{thm}[Convergence of equality constrained optimization dynamics \eqref{eq:equality z_kth_lagrange_derivative_solution}] \label{Thm: Eq}
Let Assumptions \ref{ass:smooth}, \ref{ass:convexity}, \ref{ass:MFCQ} and \ref{ass:lipschitz_gradient} hold.
Then for any initial condition, the trajectory $t \mapsto \mathbf{z}(t)$ of system $ \mathbb{G}_{\mathrm{eq}}(\mathbf{z}^{[k]}, t) =0  $  defined in \eqref{eq:equality z_kth_lagrange_derivative_solution} globally asymptotically converges to the optimal solution of $\mathbf{z}^*(t) = \mathrm{col} (\mathbf{y}^*(t), \boldsymbol{\nu}^*(t))$ of the time-varying equality constrained optimization problem \eqref{eq:Eq constrained TVO}. Moreover, the estimate
\begin{align*}
    & \| \mathbf{z}(t)-\mathbf{z}^*(t)\|_2 \leq Ce^{-\alpha t}
\end{align*}
holds, for all $t\ge 0$, where 
\[
0 < C = \left(\tfrac{c^2}{K^2} \sum_{j=0}^{k-1}\nolimits \|{\nabla_\mathbf{z}^{(j)} {L}(\mathbf{z}(0),0)}\|_2^2)\right)^{\frac{1}{2}} < \infty,\nonumber 
\]
for some constant $c > 0$, where $-\alpha := \max_{\lambda\in\spec(\mathbf{H})} \Re[\lambda]+ \epsilon_H $, for some  $ \epsilon_H >0$ small enough{, with $K$ defined in Lemma~\ref{cor: uniform boundedness of inverse KKT}.}  \\

\begin{proof}
 See Appendix~\ref{Appendix: Eq}.\end{proof}
\end{thm}

Lastly, as in Section \ref{sec:Unconstrained_TV_optimization_Framework}, we must simultaneously solve the  equations $ \mathbb{F} (\mathbf{z}^{[k]},{\mathbf{u}}):=\mathbf{u} - \alpha_z(\mathbf{z}^{[k]})=0$ and $\mathbb{G}_{\mathrm{eq}}(\mathbf{z}^{[k]}, t)  =0 $
for the pair $ (\mathbf{z}^{(k)},  \mathbf{u} ) = S(\mathbf{z}^{[k-1]},t) $, where $\alpha_z(\mathbf{z}^{[k]}) :=\alpha(\mathbf{y}^{[k]}) $. According to Lemma \ref{cor: uniform boundedness of inverse KKT}, the KKT matrix is uniformly well-defined and bounded; this allows us to first solve for $\mathbf{z}^{(k)}$ using \eqref{eq:equality z_kth_lagrange_derivative_solution} and subsequently solve for $ \mathbf{u}$ using $\eqref{eq:input-output relationship_flat_system}$. The following result summarizes these findings.

\begin{thm}[Equality constrained TVO-based control for system \eqref{eq:system}]\label{Thm: Eq_Control}
Let Assumptions \ref{ass:smooth}, \ref{ass:convexity}, \ref{ass:MFCQ} and \ref{ass:lipschitz_gradient}  hold and consider the differentially flat system  \eqref{eq:system} with the feedback controller
\[
\mathbf{u} = \alpha_z(\mathbf{z},\dots, \mathbf{z}^{(k-1)}, g_{\mathrm{eq}}(\mathbf{z}^{[k-1]} ))
\]
where
\begin{align*} 
   &g_{\mathrm{eq}}(\mathbf{z}^{[k-1]} )
   :=-\nabla_{\mathbf{zz}}^{-1}{L}(\mathbf{z},t)\left[ \sum_{i = 0} ^{k-1} a_i\nabla_{\mathbf{z}}^{(i)}{L}(\mathbf{z},t) \right.\\
  & \quad +  \nabla_{\mathbf{z}t}^{(k-1)}{L}(\mathbf{z},t)  
  + \left. \sum ^{k-1} _{m=1}\binom{k-1}{m} {\nabla_{\mathbf{zz}}^{(m)}}{L}(\mathbf{z},t)\mathbf{z}^{(k-m)}  \right] .
\end{align*}
Then, for any initial condition, the flat output of \eqref{eq:system} globally asymptotically converges to the optimal solution $\mathbf{y}^*(t)$ of \eqref{eq:Eq constrained TVO}.
\end{thm}
{\begin{proof} The proof is immediate from the preceding construction, with the KKT matrix in place of the Hessian (uniformly invertible by Lemma \ref{cor: uniform boundedness of inverse KKT}); convergence follows from Theorem~\ref{Thm: Eq}. \end{proof}}

\section{Inequality Constrained Time-varying Optimization Framework}\label{sec:Ineq Constrained TVO Framework}
In Section \ref{sec:Equality Constrained TVO Framework}, we showed how to incorporate equality constraints by Lagrangian duality in our framework. In this section, we now consider the time-varying inequality-constrained optimization problem
\begin{equation}\label{eq:inequality_constrained_time_varying_optimization}
    \begin{aligned}
     \mathbf{y}^*(t) = \arg \min_{\mathbf{y} \in \eucd^m} &f_0(\mathbf{y},t)  \\ 
       \textrm{s.t.} \quad & f_i(\mathbf{y},t) \leq 0, \quad i\in [p]. 
\end{aligned}
\end{equation}
{For the inequality-constrained formulation, we additionally impose the following local bounds and strict-feasibility condition.}
{\begin{ass}[Local gradient bounds and strict feasibility] \label{ass:lipshictz_continuity}
For every $R>0$, there exist constants $M_i(R)>0$ such that
\[
\|\nabla_{\mathbf y}f_i(\mathbf y,t)\|_2\leq M_i(R),\quad i=0,\ldots,p,
\]
for all $\|\mathbf y\|_2\leq R$ and $t\geq0$. Moreover, there exist $\mathbf y_{\rm s}(t)$ and constants $R_{\rm s},\rho>0$ such that, for all $t\geq0$,
\[
\|\mathbf y_{\rm s}(t)\|_2\leq R_{\rm s},\quad
f_i(\mathbf y_{\rm s}(t),t)\leq-\rho,\quad i\in[p].
\]
\end{ass}}
{\noindent For convex $f_i$, Assumption~\ref{ass:lipshictz_continuity} implies part 1) of Assumption~\ref{ass:MFCQ} ($\epsilon=\rho$, $\Bar{\mathbf d}=\mathbf y_{\rm s}-\mathbf y^*$), which we keep in this section for the tightness of its constants.}

Define the Lagrangian $L: \eucd^m \times \eucd^p \times \eucd_+ \to \eucd$ associated with the problem \eqref{eq:inequality_constrained_time_varying_optimization} as 
\begin{equation*}
    L(\mathbf{y},\boldsymbol{\lambda},t) = f_0(\mathbf{y},t) +\sum _{i = 1}^p {\lambda}_i f_i(\mathbf{y},t)
\end{equation*}
where ${\lambda}_i$ is the Lagrange multiplier associated with the $i$th inequality constraint $f_i(\mathbf{y},t) \leq 0$. {Under Assumptions~\ref{ass:smooth}, \ref{ass:convexity}, and~\ref{ass:MFCQ}}, the time-varying optimization problem is uniformly strongly convex, and the KKT conditions are necessary and sufficient for optimality \cite{boyd2004convex, borwein2010convex}.  Precisely, for any $t \geq 0$ we have the following KKT conditions
\begin{equation} \label{eq: KKT for inequality constraints}
    \begin{aligned}
        &{\nabla_{\mathbf{y}} f_0(\mathbf{y}^*(t),t)\! + \!\sum _{i = 1}^p {\lambda}^*_i(t)  \nabla_{\mathbf{y}}f_i(\mathbf{y}^*(t),t) =0} \\
    &{\lambda}^*_i(t) \geq 0, \quad {{\lambda}^*_i(t)f_i(\mathbf{y}^*(t),t) = 0}, \quad i \in [p].
    \end{aligned}
\end{equation}
{Primal feasibility, $f_i(\mathbf{y}^*(t),t) \leq 0$ for all $i\in[p]$, holds by definition of $\mathbf{y}^*(t)$ in \eqref{eq:inequality_constrained_time_varying_optimization}.}  

Motivated by \cite{fazlyab2017prediction,boyd2004convex}, we will use an interior-point algorithm -- the barrier method -- to enforce the inequality constraints in \eqref{eq:inequality_constrained_time_varying_optimization}. The goal of the barrier method is to approximately formulate the time-varying inequality-constrained problem as a time-varying unconstrained problem. Towards this goal, we first  rewrite the problem \eqref{eq:inequality_constrained_time_varying_optimization} as
\begin{align}\label{eq: Indicator function of ineq constraint problem}
     \mathbf{y}^*(t) = &\arg \min_{\mathbf{y} \in \eucd^m}f_0(\mathbf{y},t) + \sum_{i=1}^p  \mathbb{I}_{-}(f_i(\mathbf{y},t)),
\end{align}
where $\mathbb{I}_{-}: \eucd \to \eucd \cup \{\infty\}$ is the indicator function for the nonpositive reals: $ \mathbb{I}_{-}(u) = 0$ for $ u \leq 0$ and $ \mathbb{I}_{-}(u) = +\infty $ for $ u > 0 $. The constraints have been moved to the objective function, but the objective function is now extended-real valued and non-differentiable. 
{We therefore replace the indicator by a smooth barrier function that penalizes proximity to the constraint boundary. The classical choice is the logarithmic barrier $-\frac{1}{c(t)}\log(-u)$ \cite{boyd2004convex,fazlyab2017prediction}, which, however, is defined only on the feasible region and demands unbounded control effort near its boundary. This is an obstruction for implementations, where saturation, disturbances, or sampling can push the state toward the boundary. 

Instead, we adopt a \emph{relaxed} logarithmic barrier, in the spirit of relaxed-barrier model predictive control \cite{feller2017relaxed}: fix an integer $m_r \ge k+1$ and a threshold {$\delta > 0$}, and define}  
\begin{align}\label{eq:relaxed barrier}
B_\delta(\sigma) :=
\begin{cases}
-\log \sigma, & \sigma \ge \delta,\\
-\log\delta + \sum_{j=1}^{m_r} \frac{1}{j}\big(\tfrac{\delta - \sigma}{\delta}\big)^{j}, & \sigma < \delta,
\end{cases}
\end{align}
\begin{figure}[t]
\centering
\includegraphics[width=\columnwidth]{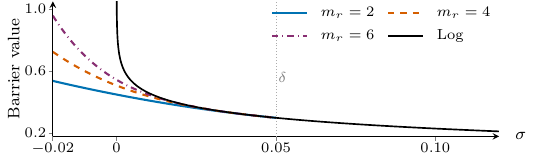}
\setlength{\abovecaptionskip}{0pt}
\caption{{Scaled logarithmic barrier $-0.1\log\sigma$ and relaxations $0.1B_{0.05}(\sigma)$ for $m_r=2,4,6$, coincident for $\sigma\geq0.05$.}}
\label{fig:relaxed-barrier}\vspace{-2ex}
\end{figure}
{i.e., $-\log$ is kept above the threshold and replaced below it by its degree-$m_r$ Taylor expansion at $\sigma = \delta$ {(Fig.~\ref{fig:relaxed-barrier})}. The choice $m_r \ge k+1$ ensures that $B_{\delta}(\sigma)$ has $k+1$ continuous derivatives, as used by the controller.    

{Define the relaxed potential and its minimizer by}}
\begin{equation}\label{eq:Phi family}
\begin{aligned}
\Phi_\delta(\mathbf{y},t) &:= f_0(\mathbf{y},t) + \frac{1}{c(t)}\sum_{i=1}^p B_{\delta(t)}\big(s(t)-f_i(\mathbf{y},t)\big),\\
\hat{\mathbf{y}}^*_\delta(t) &:= \arg\min_{\mathbf{y} \in {\real^m}} \Phi_\delta(\mathbf{y},t).
\end{aligned}
\end{equation}
{\noindent Here, {$\delta(t)>0$}, $c(t)>0$, and $s(t)\ge 0$ are time-dependent parameters with distinct roles. {The threshold $\delta(t)$ sets where the relaxation begins.} The coefficient $c(t)$ controls the accuracy of the approximation and is required to be monotonically increasing, bounded for finite times, and asymptotically diverging. The slack $s(t)$ accommodates initializations that violate the constraints, starting large enough that $f_i(\mathbf{y}(0),0) < s(0)$ for all $i \in [p]$, and decreasing to zero to recover the original constraints. We make the selections}
\begin{equation}\label{eq:c(t)}
    c(t) = c_0 e^{\alpha_c t}, \;\; s(t) = s_0 e^{-\alpha_s t}, \;\; \alpha_c, \alpha_s, c_0 > 0, \;\text{with}
\end{equation}
\begin{equation}
    \begin{aligned} \label{eq:slack_variable_initial_condition}
    s_{0}\!=\! \begin{cases} 
      0 & {\mathbf{if} \max_i f_i(\mathbf{y}(0),0) \!<\! 0}  \\
      \max_i f_i(\mathbf{y}(0),0)+ \epsilon_s & {\mathbf{if} \max_i f_i(\mathbf{y}(0),0) \!\geq\! 0} \\
   \end{cases} 
\end{aligned}
\end{equation}

{\noindent for some $\epsilon_s > 0$. The relaxed potential has the following properties.

\begin{prop}[Properties of $\Phi_\delta$]\label{prop:Phi properties}
Let Assumptions~\ref{ass:convexity} and~\ref{ass:lipshictz_continuity} hold, let $0<\delta(t)<\rho$, and define
\begin{align*}
K_{\rm s}&:=M_0(R_{\rm s})+\frac{1}{c_0\rho}\sum_{i=1}^p M_i(R_{\rm s}),\\
R&:=R_{\rm s}+m_f^{-1}\max\{2M_0(R_{\rm s}),K_{\rm s}\}.
\end{align*}
Then, for all $t\geq0$:
\begin{enumerate}[label=(\roman*),nosep,wide=0pt,labelindent=0pt]
\item $\Phi_\delta(\cdot,t)$ is $m_f$-strongly convex and $m_r$-times continuously differentiable on $\real^m$, and $\|\nabla_{\mathbf{yy}}^{-1}\Phi_\delta(\mathbf y,t)\|_2\leq 1/m_f$ for all $\mathbf y$.
\item $\hat{\mathbf y}^*_\delta(t)$ is well defined, and $\|\mathbf y^*(t)\|_2,\|\hat{\mathbf y}^*_\delta(t)\|_2\leq R$, so the gradients $\nabla_{\mathbf y}f_i$ at both minimizers are bounded by $M_i(R)$.
\end{enumerate}
\end{prop}
\begin{proof} See Appendix~\ref{appendix:relaxed barrier}. \end{proof}
{As a point of comparison, let $\Phi_0$ denote \eqref{eq:Phi family} with $B_0(\sigma):=-\log\sigma$ in place of $B_{\delta(t)}$, and let $\hat{\mathbf y}^*_0(t)$ be its minimizer over $\{\mathbf y:f_i(\mathbf y,t)<s(t),\ i\in[p]\}$. Since $B_\delta$ coincides with the logarithm for $\sigma\geq\delta$, one expects that, for $\delta$ small enough, $\Phi_0$ and $\Phi_\delta$ share the same minimizer. This in turn allows us to use the approximation bounds available for $\Phi_0$ \cite[Lemma 1]{fazlyab2017prediction}.}    The next lemma shows that, for sufficiently small $\delta(t)$, the relaxed problem has \emph{exactly} {the same minimizer at every time $t$ as $\Phi_0(\cdot,t)$}.   We refer to this property as the \emph{exactness} of the $\delta$-relaxation.
}

\begin{lem}[Exactness and approximation error]\label{lem:sub-optimality_gap_perturbed_approximation}\label{lem:relaxed barrier}
{{Let Assumptions \ref{ass:smooth}, \ref{ass:convexity}, \ref{ass:MFCQ}, and \ref{ass:lipshictz_continuity} hold.} Then there exists $\Lambda > 0$, given explicitly in \eqref{eq:Lambda}, independent of $c(t)$ and $\delta(t)$, such that:
\begin{enumerate}[label=(\roman*),nosep,wide=0pt,labelindent=0pt]
\item whenever {$0<\delta(t)<\min\{\rho,1/(\Lambda c(t))\}$}, \[\hat{\mathbf{y}}^*_\delta(t) = \hat{\mathbf{y}}^*_0(t),\quad \forall t \ge 0\,;\]
\item for any $\delta(t)$ as in (i), and for all $t \geq 0$,
\begin{align}
 | f_0(\hat{\mathbf{y}}^*_\delta(t),t) - f_0(\mathbf{y}^*(t),t) |  \leq \frac{p}{c(t)} + \sum_{i = 1}^p \lambda_i^*(t)s(t), \label{eq:sub-optimality_gap_perturbed_approximation}
\end{align}
where $\boldsymbol{\lambda}^*(t)$ is the Lagrange multiplier associated with the inequality constraints of \eqref{eq:inequality_constrained_time_varying_optimization}.
\end{enumerate}}
\end{lem}

\begin{proof}
{(i) See Appendix~\ref{appendix:exactness}. {(ii) Apply \cite[Lemma 1]{fazlyab2017prediction} to $\Phi_0$ and use (i).}}
\end{proof}
{Since the relaxed and classical minimizers coincide under the threshold in Lemma~\ref{lem:sub-optimality_gap_perturbed_approximation}(i), the approximation error relative to the original problem \eqref{eq:inequality_constrained_time_varying_optimization} is not affected by the $\delta$-relaxation and, as shown in the remainder of this section, vanishes asymptotically under the schedules \eqref{eq:c(t)}.}

{We call $\delta(t)$ \emph{admissible} if it is {$k+1$ times continuously differentiable and} $0<\delta(t)<\min\{\rho,1/(\Lambda c(t))\}$ for all $t\geq0$.}
{Only the upper bound on $\delta(t)$ is used below, however, the rate matters for the control effort since on the relaxed branch the barrier gradient in \eqref{eq:Phi family} is bounded by $m_r/(c(t)\delta(t))$. We therefore take $\delta(t)=\delta_0/c(t)$ with $\delta_0<\min\{\rho c_0,1/\Lambda\}$, as in Section~\ref{sec:Numerical}.}

{By Lemma \ref{lem:sub-optimality_gap_perturbed_approximation}(ii), \emph{if} the multipliers $\boldsymbol{\lambda}^*(t)$ remain uniformly bounded, then under the schedules \eqref{eq:c(t)} the bound \eqref{eq:sub-optimality_gap_perturbed_approximation} vanishes and $\hat{\mathbf{y}}^*_\delta(t)$ converges asymptotically to $\mathbf{y}^*(t)$. Prior work assumes such boundedness \cite{fazlyab2017prediction, simonetto2016class}. A contribution of this paper is to establish it from explicit regularity conditions (cf.\ \cite{gauvin1977necessary} for the static case), which is the content of the next lemma.}

\begin{lem}(Uniform boundedness of Lagrange multipliers) \label{lem:Unifrom_boundedness_Lagrange_Constraints}
Let $\mathbf{y}^*(t)$ be the optimal solution of \eqref{eq:inequality_constrained_time_varying_optimization}, and  suppose that Assumptions \ref{ass:smooth}, \ref{ass:convexity}, {\ref{ass:MFCQ}, and \ref{ass:lipshictz_continuity}} are satisfied. Then the set of Lagrange multipliers $\boldsymbol{\lambda}^*(t) \in \eucd^p $ satisfying the KKT conditions \eqref{eq: KKT for inequality constraints} is nonempty and uniformly bounded, i.e.,
\begin{align*}
&\| \boldsymbol{\lambda}^*(t)\|_1  \leq  {\frac{M_0(R)d}{\epsilon}} , \qquad t\geq0 ,
\end{align*}
where $\| \cdot \|_1  $ denotes the $l_1$ vector norm, $d,\epsilon$ are defined in Assumption \ref{ass:MFCQ}, 
{and $M_0(R)$ bounds the objective gradient on the ball of radius $R$ of Proposition~\ref{prop:Phi properties}.}  

\begin{proof}
 See Appendix~\ref{Appendix: Uniform_Boundedness_Multiplier}.\end{proof}
\end{lem}

Lemmas \ref{lem:sub-optimality_gap_perturbed_approximation} and \ref{lem:Unifrom_boundedness_Lagrange_Constraints} provide the necessary ingredients to ensure that the approximation error \eqref{eq:sub-optimality_gap_perturbed_approximation} asymptotically goes to zero. 

{\begin{rem}[Interpretation and tuning]
{For admissible $\delta(t)$, relaxation preserves the minimizer trajectory and asymptotic guarantees of the ideal controller. With saturated actuation, the prescribed optimization dynamics need not hold; the transient violations and subsequent recovery in Section~\ref{sec:Numerical} are empirical observations.}  

\end{rem}
}

Having ensured the asymptotic accuracy of the approximation, we now consider{, for admissible $\delta(t)$,} the following target system as a natural extension of \eqref{eq:K-th_order_Unconstrained_Optimality_Error} when inequality constraints are included
\begin{align}
    \begin{bmatrix}
       \dot{\nabla}_\mathbf{y} \Phi_\delta(\mathbf{y},t) \\ \vdots \\ \nabla^{(k)}_\mathbf{y} \Phi_\delta(\mathbf{y},t)
    \end{bmatrix} = \mathbf{H}
    \begin{bmatrix}
       \nabla_\mathbf{y} \Phi_\delta(\mathbf{y},t) \\ \vdots \\ \nabla^{(k-1)}_\mathbf{y} \Phi_\delta(y,t)
    \end{bmatrix}   , \label{eq:barrier_function_Dynamical_System}
\end{align} where $\mathbf{H} = \hat{\mathbf{H}} \otimes \mathbf{I}_m$ being Hurwitz.

Analogously, combining \eqref{eq:barrier_function_Dynamical_System} and Lemma \ref{lem:Diffrentiating_Gradient_K_Times}, we can use the following implicit function to define the \textit{optimization dynamics}, when inequality constraints are included: 
\begin{equation}
    \begin{aligned}\label{eq:ineq_design_reqirement}
        \mathbb{G}_{\mathrm{ ineq}}(\mathbf{y}^{[k]}, t) := \sum_{m=0}^{k-1}\binom{k-1}{m} \nabla_{\mathbf{yy}}^{(m)}\Phi_\delta(\mathbf{y},t)\mathbf{y}^{(k-m)} \\ +  \nabla_{\mathbf{y}t}^{(k-1)}\Phi_\delta(\mathbf{y},t) 
+ \sum_{i = 0} ^{k-1} a_i\nabla_{\mathbf{y}}^{(i)}\Phi_\delta(\mathbf{y},t) = 0. 
    \end{aligned}
\end{equation}

\begin{thm}[Convergence of inequality constrained optimization dynamics \eqref{eq:ineq_design_reqirement}] \label{Thm: Ineq}
{Let Assumptions \ref{ass:smooth}, \ref{ass:convexity}, \ref{ass:MFCQ}, and \ref{ass:lipshictz_continuity} hold}, with $c(t)$ and $s(t)$ as in \eqref{eq:c(t)}, $s_0 $ as in \eqref{eq:slack_variable_initial_condition}, {and $\delta(t)$ admissible}. Then, {for any initial condition of \eqref{eq:barrier_function_Dynamical_System}}, the trajectory $t \mapsto \mathbf{y}(t)$ of $ \mathbb{G}_{\mathrm{ineq}}(\mathbf{y}^{[k]}, t) =0  $ defined in \eqref{eq:ineq_design_reqirement} globally asymptotically converges to the optimal solution $\mathbf{y}^*(t)$ of \eqref{eq:inequality_constrained_time_varying_optimization}. Moreover, for all $t \ge 0$,
\begin{align*}
\norm{{\mathbf{y}(t)-\hat{\mathbf{y}}^*_\delta(t)}}_2  &\leq C e^{-\alpha t}, \\*
| f_0(\mathbf{y}(t),t) - f_0(\mathbf{y}^*(t),t) | &\leq C' e^{-\beta t},
\end{align*}
{where $-\alpha := \max_{\lambda\in\spec(\mathbf{H})} \Re[\lambda]+ \epsilon_H$ for some $\epsilon_H > 0$ small enough, $\beta := \min\{\alpha, \alpha_c, \alpha_s\}$, and $C, C' > 0$ are constants given explicitly in Appendix~\ref{Appendix: Ineq}.}
\end{thm}

\begin{proof}See Appendix~\ref{Appendix: Ineq}.\end{proof}

As in the preceding sections, the pair $(\mathbf{y}^{(k)}, \mathbf{u})$ is obtained by resolving $ \mathbb{F} (\mathbf{y}^{[k]},{\mathbf{u}})=0$ and $\mathbb{G}_{\mathrm{ineq}}(\mathbf{y}^{[k]}, t) =0 $ simultaneously: since $\norm{\nabla_{\mathbf{yy}}^{-1} \Phi_\delta(\mathbf{y},t)}_2  \leq m_f^{-1}$ {globally} (Proposition~\ref{prop:Phi properties}), one solves first for $\mathbf{y}^{(k)}$ using \eqref{eq:ineq_design_reqirement} and then for $ \mathbf{u}$ using $\eqref{eq:input-output relationship_flat_system}$, leading to the following result. 

\begin{thm}[Inequality constrained TVO-based control for system \eqref{eq:system}]\label{Thm: Ineq_Control}
{Let Assumptions \ref{ass:smooth}, \ref{ass:convexity}, \ref{ass:MFCQ}, and \ref{ass:lipshictz_continuity} hold}, with $c(t)$ and $s(t)$ as in \eqref{eq:c(t)}, $s_0 $ as in \eqref{eq:slack_variable_initial_condition}, {and $\delta(t)$ admissible}. Consider the differentially flat system  \eqref{eq:system} with the feedback controller
\[
\mathbf{u} := \alpha(\mathbf{y},\dots, \mathbf{y}^{(k-1)}, g_{\mathrm{ineq}}(\mathbf{y}^{[k-1]} ))
\]
where
\begin{align*}
   &g_{\mathrm{ineq}}(\mathbf{y}^{[k-1]} )  
   :=-\nabla_{\mathbf{yy}}^{-1}\Phi_\delta(\mathbf{y},t)\left[ \sum_{i = 0} ^{k-1} a_i\nabla_{\mathbf{y}}^{(i)}\Phi_\delta(\mathbf{y},t) \right.\\
  & \quad +  \nabla_{\mathbf{y}t}^{(k-1)}\Phi_\delta(\mathbf{y},t) 
  + \left. \sum_{m=1}^{k-1}{\textstyle{\binom{k-1}{m}}  }\nabla_{\mathbf{yy}}^{(m)}\Phi_\delta(\mathbf{y},t)\mathbf{y}^{(k-m)}  \right].
\end{align*} 
Then, for any initial condition, the flat output of 
\eqref{eq:system} globally asymptotically converges to the optimal solution $\mathbf{y}^*(t)$ of \eqref{eq:inequality_constrained_time_varying_optimization}.
\end{thm}
{\begin{proof} Immediate from the preceding construction, with $\Phi_\delta$ in place of $f_0$ ($\nabla_{\mathbf{yy}}\Phi_\delta \succeq m_f\mathbf{I}$, Proposition~\ref{prop:Phi properties}). Convergence follows from Theorem \ref{Thm: Ineq}. \end{proof}}

\section{Numerical Examples}\label{sec:Numerical}
In this section, we use two numerical examples arising in multi-robot coordination to illustrate the effectiveness of our solution approach. As our time-varying feedback optimization framework automatically guarantees asymptotic satisfaction of time-varying equality and inequality constraints, we apply the method for the specification of formation constraints (Section \ref{subsec:multi formation}) and to enforce collision avoidance (Section \ref{subsec: obstacal avoidance}).
 
{\noindent In all experiments below, the initial conditions are strictly feasible. Hence $s_0 = 0$ in \eqref{eq:slack_variable_initial_condition} and the slack mechanism of \eqref{eq:Phi family} is not exercised. Infeasible initializations would instead be absorbed by the exponentially shrinking slack $s(t) = s_0 e^{-\alpha_s t}$, with the true constraints recovered at rate $\alpha_s$.}

\subsection{Multi-robot navigation with formation constraints}\label{subsec:multi formation}
In this numerical example, two WMRs \eqref{eq:Unicycle_System_Dynamic} are required to track two moving objects respectively, {while keeping their mutual distance within prescribed bounds} (e.g., due to communication or formation constraints). Let $\mathbf{y}_1(t), \mathbf{y}_2(t) \in \eucd^2$ denote the position of each WMR, with $\mathbf{y}_1^d(t),  \mathbf{y}_2^d(t)$ denoting the positions of the two moving objects. To model the above objectives, consider the time-varying optimization problem
\begin{equation}
    \begin{aligned} \label{eq:Simultion1_TV}
    \min_{\mathbf{y}_1,\mathbf{y}_2} &\| \mathbf{y}_1\!-\!\mathbf{y}_1^d(t)\|^2_2\!+\! \| \mathbf{y}_2\!-\!\mathbf{y}_2^d(t)\|_2^2 \\
    s.t. \;\;& {\underline{d}(t)^2 \leq \| \mathbf{y}_1\!-\!\mathbf{y}_2\|_2^2 \leq \bar{d}(t)^2,}
\end{aligned}
\end{equation}
where {$\bar{d}(t)$ and $\underline{d}(t)$ denote the maximum and minimum (Euclidean) separation allowed between the two robots at time $t$. The squared form makes the constraint functions smooth.} 
{The trajectories of the moving objects are parametrized as $\mathbf{y}_i^d(t) = \sum_{j=1}^{N} \mathbf{A}_{ij}t^j$, $i \in \{1,2\}$ (Section 2.4 \cite{martin2003flat}), with coefficients $\mathbf{A}_{ij}$ determined by the initialization.}

\begin{figure*}[!t]
\centering
\includegraphics[width=\textwidth]{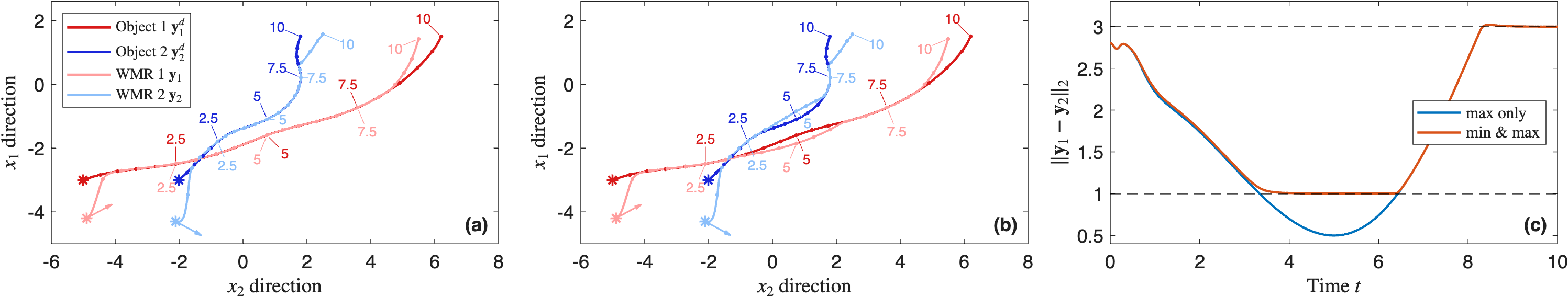}
\caption{Multi-robot formation experiments, with object trajectories that approach, cross, and separate. (a) Trajectories of the two moving objects $\mathbf{y}_1^d(t), \mathbf{y}_2^d(t)$ (dark) and the two WMRs $\mathbf{y}_1, \mathbf{y}_2$ (light) under the maximum-distance constraint $\|\mathbf{y}_1\!-\!\mathbf{y}_2\|_2 \leq 3$ only. Arrows indicate the initial velocities, and the small numbers indicate the time $t$ along each trajectory. (b) The same experiment with the bilateral constraints $1 \le \|\mathbf{y}_1\!-\!\mathbf{y}_2\|_2 \le 3$. (c) Inter-robot distance $\|\mathbf{y}_1\!-\!\mathbf{y}_2\|_2$ for both experiments. The dashed lines are the constraint bounds. Without the minimum-distance constraint the robots approach within $0.5$ of each other while the targets cross (blue), whereas the bilateral controller rides the minimum-distance boundary (orange).}
\label{fig:formation}
\end{figure*}

{The object trajectories start at $(-5, -3, 0.5)$ and $(-2, -3, 0.5)$ and are designed to approach, cross, and then separate to a final distance of about $4.4$. The WMRs start near their respective objects, at $(-4.9, -4.2, 0.5)$ and $(-2.1, -4.3, -0.5)$. In all runs $\bar{d}(t) \equiv 3$, the horizon is $T = 10$ s, actuation is bounded ($|u_2| \le 2\pi$ rad/s, $|\dot u_1| \le 15$), and the controller uses the relaxed barrier of Section \ref{sec:Ineq Constrained TVO Framework} with $\delta(t) \propto 1/c(t)$, $c_0 = 1$, and the coefficients of \eqref{eq:ineq_design_reqirement} (here $k = 2$) set to $(a_0, a_1) := (k_p, k_d) = (40, 13)$, proportional and derivative gains chosen commensurate with the actuation limits. The closed-loop dynamics are integrated with MATLAB's \texttt{ode45}. We report two experiments. The first enforces only the maximum-distance constraint ($\underline{d} = 0$). The second adds the minimum-distance constraint $\underline{d} = 1$, with all remaining parameters unchanged, so that the two runs are directly comparable. {The minimum-distance constraint is nonconvex, so the second experiment illustrates an empirical extension beyond our convexity assumptions.}

Figure \ref{fig:formation}(a) shows the first experiment: the robots track their targets through the crossing, approaching within $0.5$ of each other, and the maximum-distance constraint becomes active as the targets separate. Figure \ref{fig:formation}(b) shows the second: the closed loop now rides the minimum-distance boundary while the targets cross, and the maximum-distance boundary as they separate. Figure \ref{fig:formation}(c) compares the inter-robot distance of the two runs against the bounds. {Under saturated actuation, the distance briefly overshoots the boundary (by less than $1\%$) before returning close to it. This recovery is empirical; Lemma~\ref{lem:relaxed barrier} concerns barrier minimizers, not saturated trajectories.}  Note also that the maximum-distance constraint is \emph{active} at the optimizer (the targets end up $4.4$ apart), which is why the tracked distance approaches the bound from below as $c(t)$ grows. {Assumption \ref{ass:MFCQ} holds in both experiments. At most one distance constraint is active. Its gradient $\mathbf{g}(t)$ with respect to $(\mathbf{y}_1,\mathbf{y}_2)$, evaluated at the optimal solution, satisfies $\|\mathbf{g}(t)\|_2=2\sqrt{2}\|\mathbf{y}_1^*(t)-\mathbf{y}_2^*(t)\|_2\geq 2\sqrt{2}$. Choosing $\Bar{\mathbf{d}}(t)=-\mathbf{g}(t)/\|\mathbf{g}(t)\|_2$ gives $\|\Bar{\mathbf{d}}(t)\|_2=1$ and $\mathbf{g}(t)^{\T}\Bar{\mathbf{d}}(t)\leq-2\sqrt{2}$. When no constraint is active, take $\Bar{\mathbf{d}}(t)=0$. There are no equality constraints.}   
}

\subsection{Robot tracking and obstacle avoidance}\label{subsec: obstacal avoidance}

\begin{figure*}[!t]
\centering
\includegraphics[width=\textwidth]{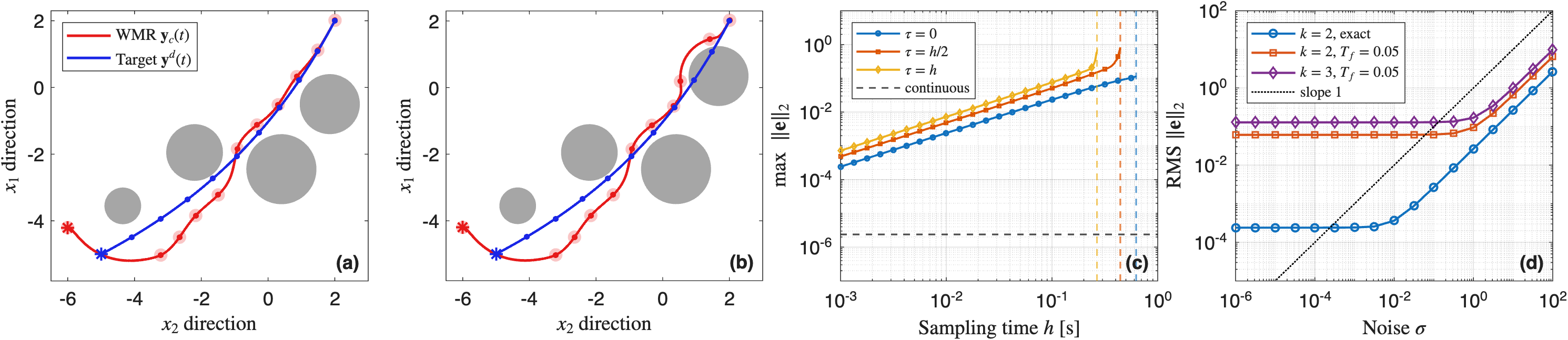}
\caption{Obstacle avoidance and practical considerations. (a) WMR (red) tracking a moving target (blue) while avoiding the obstacles (grey disks). The shaded disks show the robot footprint every second, starting at the red asterisk. (b) The same experiment with an \emph{infeasible} target that crosses an obstacle: the robot detours around it and re-acquires the target on the far side. (c) Worst-case tracking error of a zero-order-hold implementation {of the integrator-chain testbed of Section \ref{subsec:practical}} versus the sampling time $h$. The dashed line is the continuous-time baseline.{The input applied at each sampling instant is computed from the flat output measured $\tau$ seconds earlier, for computation times $\tau = 0$, $h/2$, and $h$. Each curve is shown up to its stability boundary.} (d) Steady-state RMS tracking error versus the measurement-noise level $\sigma$: $k = 2$ with exact derivatives, and $k = 2, 3$ with derivatives reconstructed by dirty-derivative filters (time constant $T_f$, cascaded for $k = 3$).}
\label{fig:obstacles}
\end{figure*}
In this section, we aim to solve the problem of navigating a disk-shaped wheeled mobile robot (WMR) to track a moving target without colliding with spherical obstacles in the environment. Our construction follows the concept of a {local free space} with \textit{projected goals}\cite{arslan2016exact}, and particularly \cite{fazlyab2017prediction}, wherein a robot navigation problem (for integrator dynamics~\eqref{eq:Integrator_System_Dynamic}) is formulated as time-varying convex optimization problem. We first introduce some preliminary concepts and then show how to generalize the results of \cite{fazlyab2017prediction} to our framework.

Consider a closed and convex \emph{workspace} $ \mathcal{W} \subset \eucd^2$, which is populated with $O$ non-intersecting spherical obstacles, where the center and radius of the $i$th obstacle are denoted by ${\mathbf{o}_i} \in \mathcal{W}$ and $r_i > 0$, respectively.   Suppose the wheeled mobile robot (WMR) of radius $r>0$ is defined as in \eqref{eq:Unicycle_System_Dynamic}, where the flat output, namely the position vector of the center of mass of the WMR, is given by $ \mathbf{y}_c = (x_1,x_2)$. We define the \textit{free space}, denoted by $ \mathcal{F}$, as the set of configurations in the workspace in which the robot does not collide with any obstacle, i.e.,
\begin{align}
    \mathcal{F} =\{\mathbf{y} \in \mathcal{W} : \bar{B}(\mathbf{y},r ) \subseteq \mathcal{W} \setminus \cup_{i=1}^O B({\mathbf{o}_i},r_i )    \}
\end{align}
where $B(\mathbf{y},r )$ is the $2$-dimensional open ball centered at $\mathbf{y} $ with radius $r$, and $\bar{B}(\mathbf{y},r )$ denotes its closure. Given the moving target $\mathbf{y}^d(t) \in \mathcal{F}$ for all $ t \geq 0$, we aim to solve for the control input $ \mathbf{u}(t) $ such that $\mathbf{y}_c (t) \in   \mathcal{F}  $ for all $ t\geq 0$ with initialization $\mathbf{y}_c (0) \in  \mathcal{F}$. Moreover, we wish to achieve $\lim_{t \to \infty} \mathbf{y}_c (t) =  \mathbf{y}^d(t) $ if the obstacles were to allow it, i.e., global asymptotic convergence of the WMRs towards obstacle free targets.

Define the \textit{local workspace}  as in \cite{arslan2016exact,fazlyab2017prediction},
\begin{align*}
      \mathcal{LW}(\mathbf{y}_c) = \{\mathbf{y} \in \mathcal{W} \!:\| \mathbf{y}-\mathbf{y}_c \|_2^2 - r^2  \leq \| \mathbf{y}-{\mathbf{o}_i} \|_2^2 - r_i^2 ,\forall i\}.
\end{align*}
{According to \cite[Prop.~1]{arslan2016exact}, $\mathbf{y}_c\in\mathcal{F}$ if and only if $\bar{B}(\mathbf{y}_c,r)\subseteq\mathcal{LW}(\mathbf{y}_c)$. To obtain a collision-free set of candidate robot-center positions, define the \emph{local free space} as
\begin{align}
\mathcal{LF}(\mathbf{y}_c)
:=
\{\mathbf{y}\in\mathcal{LW}(\mathbf{y}_c):
\bar{B}(\mathbf{y},r)\subseteq\mathcal{LW}(\mathbf{y}_c)\}.
\end{align}}

{Whenever the robot has positive clearance, $\mathbf{y}_c$ lies in the interior of $\mathcal{LF}(\mathbf{y}_c)$. If the minimizer $\mathbf{y}^*(t)$ lies on its boundary, the direction $\mathbf{y}_c-\mathbf{y}^*(t)$ strictly decreases every active constraint; if it lies in the interior, there are no active constraints. When the clearance remains bounded away from zero, these directional derivatives are uniformly bounded above by some $-\epsilon<0$, as required by Assumption \ref{ass:MFCQ}; the clearance condition held along all trajectories in our experiments.}

Following \cite{arslan2016exact}, the {local free space} $\mathcal{LF}$ is equivalent to the following set of inequality constraints:
\begin{align*}
    \mathcal{LF}(\mathbf{y}_c) =\{\mathbf{y} \in \mathcal{W} : a_i(\mathbf{y}_c)^{\T}\mathbf{y} - b_i(\mathbf{y}_c) \leq 0, i \in [O]  \},
\end{align*}
where
\begin{align*}
   & a_i(\mathbf{y}_c) = {\mathbf{o}_i} - \mathbf{y}_c, \qquad \theta_i (\mathbf{y}_c) = \frac{1}{2} - \frac{r_i^2 - r^2}{ 2 \| {\mathbf{o}_i} - \mathbf{y}_c\|^2},\\
    & b_i(\mathbf{y}_c) = ({\mathbf{o}_i} - \mathbf{y}_c)^{\T} \left(\theta_i{\mathbf{o}_i}+ ( 1 -\theta_i ) \mathbf{y}_c + r \frac{ \mathbf{y}_c - {\mathbf{o}_i} }{\|  \mathbf{y}_c - {\mathbf{o}_i}\|}\right),
\end{align*}
we refer the reader to \cite{arslan2016exact} for a detailed derivation of these terms.

Using the definition of {local free space}, \cite{arslan2016exact} and \cite{fazlyab2017prediction} find a robot navigation strategy that steers the robot $\mathbf{y} \in \mathcal{F} $ towards the global goal $ \mathbf{y}^d(t)$ through a safe local target location, called the \textit{projected goal} by solving the following optimization problem:
\begin{align*}
    \mathbf{y}^*(t) :=& \argmin_{  \mathbf{y} \in \eucd^2 } \frac{1}{2}\|\mathbf{y}  - \mathbf{y}^d(t) \|^2  \nonumber\\
    &\mathrm{s.t.} \;\; a_i(\mathbf{y}_c)^{\T}\mathbf{y} - b_i(\mathbf{y}_c) \leq 0,\;\; i \in {[O]}.
\end{align*}

In \cite{arslan2016exact} they consider the problem of tracking a static goal $\mathbf{y}^d$, and in \cite{fazlyab2017prediction} the authors extend the problem to track a moving target $\mathbf{y}^d(t)$ for a linear integrator \eqref{eq:Integrator_System_Dynamic}. Although in the case of tracking a moving target, there are no theoretical guarantees, the simulation results show the proposed method successfully tracks the moving target without colliding with obstacles in the environment. {We further extend these results by navigating a WMR \eqref{eq:Unicycle_System_Dynamic} to track a moving target, and consider two cases: a target that remains in the free space, and an \emph{infeasible} target that crosses an obstacle. The latter is meaningful because $\mathbf{y}^d(t)$ need \emph{not} avoid the obstacles: only the optimizer of the constrained time-varying problem does, which is what the robot tracks. Computing $\mathbf{y}^d$ therefore poses no planning problem of its own. In both cases the target trajectory is generated with the time-parametric representation of Section \ref{subsec:multi formation}, the WMR and target start at $(-6, -4.2, -10)$ and $(-5, -5, 0.5)$, the workspace contains four non-intersecting spherical obstacles, the WMR radius is $r = 0.2$, and the closed-loop dynamics are integrated with MATLAB's \texttt{ode45}.
Figure \ref{fig:obstacles}(a) shows the first case: the robot tracks the moving target while avoiding the obstacles. Figure \ref{fig:obstacles}(b) shows the second: the robot detours around the crossed obstacle while the target passes through, and re-acquires it on the far side.}

{\subsection{Practical considerations}\label{subsec:practical}

We close this section by quantifying three practical aspects of the proposed controller. All studies share one testbed: a chain of $k$ integrators $\mathbf{y}^{(k)} = \mathbf{u}$ with flat output $\mathbf{y} \in \eucd^2$, the simplest flat system of relative degree $k$, steered to the minimizer of $f_0(\mathbf{y},t) = \frac{1}{2}\|\mathbf{y}-\mathbf{y}^d(t)\|^2$ with $\mathbf{y}^d(t) = (\cos\omega t, \sin\omega t)$ and $\omega = 2\pi/10$. {Since $\nabla_{\mathbf{yy}} f_0 = \mathbf{I}$, Theorem~\ref{Thm: Unconstrained_Control} gives}  
\begin{align}
    \mathbf{u} = (\mathbf{y}^d)^{(k)}(t) - \sum_{i=0}^{k-1}\nolimits a_i \left(\mathbf{y}^{(i)} - (\mathbf{y}^d)^{(i)}(t)\right), \label{eq:chain_testbed}
\end{align}
with the coefficients $\{a_i\}_{i=0}^{k-1}$ selected to place all closed-loop poles at $-1.6$. {We use $k = 2$ throughout and additionally $k = 3$ in the measurement-noise study, where one more output derivative must be reconstructed.}   All runs last $T = 20$ s, start at $\mathbf{y}(0) = (2,0)$ with vanishing higher derivatives, and steady-state metrics are computed over the last target period.

\smallskip
\noindent\textbf{Computational cost.} The controller requires, per evaluation, derivatives of the problem data and one linear solve with the $n_y \times n_y$ Hessian $\nabla_{\mathbf{yy}}\Phi$, where $n_y$ is the dimension of the \emph{flat output} ($n_y = 2$ above), not of the state. This contrasts with solving the time-varying problem itself, which would require an iterative solver to convergence at every sampling instant, followed by numerical differentiation of the computed optimizers (cf.\ Remark \ref{rem:solver}). When the linear solve is a bottleneck, Neumann-series or quasi-Newton approximations of the Hessian inverse may be employed \cite{nocedal2006numerical}.

\smallskip
\noindent\textbf{Discretization.} Figure \ref{fig:obstacles}(c) reports the tracking error of a zero-order-hold implementation as a function of the sampling time $h$: the error degrades gracefully (empirically $O(h)$ over more than two decades) and the closed loop remains stable well beyond $h = 0.1$ s. {A computation time $\tau \le h$ is modeled explicitly: the flat output is measured at $t_{n+1}-\tau$, the controller computes for $\tau$ seconds, and the resulting input is applied at $t_{n+1}$ and held over the next interval. Figure \ref{fig:obstacles}(c) reports $\tau = h/2$ and $\tau = h$ alongside the ideal loop $\tau = 0$. Each case follows the same trend, and the delay shifts the error up (by factors $2.0$ and $3.0$) and shrinks the stability range: the stability edges are at $h \approx 0.69$\,s for $\tau=0$, $0.63$\,s for $\tau=h/2$, and $0.33$\,s for $\tau=h$. A fully discrete design could instead start from discrete-time notions of flatness \cite{diwold2022trajectory,kolar2023difference}.}

\smallskip
\noindent\textbf{Measurement noise.} An implementation must reconstruct $\dot{\mathbf{y}}, \dots, \mathbf{y}^{(k-1)}$ from the measured output, so the sensitivity to noise is inherently tied to the relative degree $k$. Figure \ref{fig:obstacles}(d) quantifies this. With exact derivative information, the steady-state error is proportional to the noise level $\sigma$, down to the discretization floor of the $h = 10^{-3}$ s implementation of panel (c). When the derivatives are instead reconstructed by a causal dirty-derivative filter $s/(T_f s + 1)$ (time constant $T_f$), applied once for $\dot{\mathbf{y}}$ and twice in cascade for $\ddot{\mathbf{y}}$ when $k = 3$, the filter lag introduces a bias floor that dominates over the entire range tested (the $\sim \sigma/T_f^{k-1}$ noise amplification would take over only at larger $\sigma$). At the same $T_f$, the $k = 3$ floor sits a factor of $1.8$ above the $k = 2$ one, the compounding of filter lag with the relative degree. Degradation is graceful in all cases, and no instability was observed.

}

\section{Conclusions and Future Work}\label{sec:Conclusion}
In this paper, we investigate the problem of steering in real time a differentially flat system to the minimizer of a time-varying constrained optimization problem. We develop a time-varying optimization-based framework composed of the system dynamic, which is an implicit function describing the input-output relationship, and the optimization dynamics, which serves as an online approximation of the minimizer. 
Such nonlinear control effectively transforms a differentially flat system into an optimization algorithm, which seeks to find the optimal solution to the time-varying optimization problem. Under mild assumptions, we show that the proposed control law asymptotically converges to the optimal solution of the (possibly constrained) time-varying optimization problem.  Lastly, the effectiveness of our method is illustrated in two numerical examples: a multi-robot navigation problem and an obstacle avoidance problem.
Future work includes generalizations to integral objectives, as in optimal control, accounting for the effect of discrete algorithm updates{, and establishing anytime constraint satisfaction during transients, e.g., via control-barrier-function-type conditions for time-varying constraint sets}.

\appendix
\subsection{Proof of Lemma \ref{lem:Diffrentiating_Gradient_K_Times}}\label{Appendix:Diffrentiating_Gradient_K_Times}
We prove this by mathematical induction. First, we consider when $k = 1$ and $2$.
\begin{align*}
    \dot{\nabla}_\mathbf{y} f_0(\mathbf{y},t) &= \frac{\partial \nabla_\mathbf{y} f_0(\mathbf{y},t) }{\partial \mathbf{y}}\dot{\mathbf{y}}+\frac{\partial \nabla_\mathbf{y} f_0(\mathbf{y},t) }{\partial t}  \nonumber\\
    &= \nabla_{\mathbf{yy}}f_0(\mathbf{y},t)\dot{\mathbf{y}}+\nabla_{\mathbf{y}t}f_0(\mathbf{y},t)\\
    \ddot{\nabla}_\mathbf{y} f_0(\mathbf{y},t) &= \frac{\mathrm{d}}{\mathrm{d}t}(\nabla_{\mathbf{yy}}f_0(\mathbf{y},t)\dot{\mathbf{y}}+\nabla_{\mathbf{y}t}f_0(\mathbf{y},t)) \nonumber\\
    &= \nabla_{\mathbf{yy}}f_0(\mathbf{y},t)\ddot{\mathbf{y}}+\dot{\nabla}_{\mathbf{yy}}f_0(\mathbf{y},t)\dot{\mathbf{y}}+\dot{\nabla}_{\mathbf{y}t}f_0(\mathbf{y},t)
\end{align*} 
We want to show that for every $k \geq k_0$, $k_0 \geq 2$, if the statement holds for $k$, then it holds for $k+1$. 
\begin{align*}
    \nabla_\mathbf{y}^{(k)} f_0(\mathbf{y},t) &= \sum_{m = 0}^{k-1} \binom{k-1}{m} \nabla_{\mathbf{yy}}^{(m)}f_0(\mathbf{y},t)\mathbf{y}^{(k-m)} \nonumber\\
    &+\nabla_{\mathbf{y}t}^{(k-1)}f_0(\mathbf{y},t)
\end{align*}

Using the binomial theorem we obtain:
\begin{align*}
    \nabla_\mathbf{y}^{(k+1)} f_0(\mathbf{y},t) &= \frac{\mathrm{d}}{\mathrm{d}t}(\sum_{m = 0}^{k-1} \binom{k-1}{m} \nabla_{\mathbf{yy}}^{(m)}f_0(\mathbf{y},t)\mathbf{y}^{(k-m)}) \nonumber\\
    &+\frac{\mathrm{d}}{\mathrm{d}t}(\nabla_{\mathbf{y}t}^{(k-1)}f_0(\mathbf{y},t)) \nonumber\\
    &= \sum_{m = 0}^{k} \binom{k}{m} \nabla_{\mathbf{yy}}^{(m)}f_0(\mathbf{y},t)\mathbf{y}^{(k+1-m)} \nonumber
    \\&+\nabla_{\mathbf{y}t}^{(k)}f_0(\mathbf{y},t),
\end{align*}which completes the proof.

\subsection{Proof of Theorem \ref{Thm: Unconstrained}} \label{Appendix: thm Unconstrained}

By Lemma \ref{lem:Diffrentiating_Gradient_K_Times}, the trajectory $ \mathbf{y}(t)$ of the system $ \mathbb{G}_{\mathrm{unc}}(\mathbf{y}^{[k]}, t) =0$ in  \eqref{eq:design_requirement_q_order_unconstr} satsifies the optimization dynamics in \eqref{eq:K-th_order_Unconstrained_Optimality_Error}, with $ \mathbf{H}$ being the designed Hurwitz matrix. The solution to the ODE system \eqref{eq:K-th_order_Unconstrained_Optimality_Error} is
\begin{align*}
    \begin{bmatrix}
       \nabla_\mathbf{y} f_0(\mathbf{y}(t),t) \\ \vdots \\ \nabla^{(k-1)}_\mathbf{y} f_0(\mathbf{y}(t),t)
    \end{bmatrix} &= e^{\mathbf{H}t}
    \begin{bmatrix}
       \nabla_\mathbf{y} f_0(\mathbf{y}(0),0) \\ \vdots \\ \nabla^{(k-1)}_\mathbf{y} f_0(\mathbf{y}(0),0)
    \end{bmatrix}
\end{align*}
where $\mathbf{y}(0) \in \mathbb{R}^m$ is the initial condition. By taking the Euclidean norms of both sides we obtain
\begin{align}
 \!\sum_{j=0}^{k-1} \! \norm{{\nabla_\mathbf{y}^{(j)} f_0(\mathbf{y}(t),t)}}_2^2
   \! \leq  \! c^2e^{-2\alpha t}\sum_{j=0}^{k-1} \! \norm{{\nabla_\mathbf{y}^{(j)} f_0(\mathbf{y}(0),0)}}_2^2 \! \label{eq:Left_Norm_Inequality}
\end{align}
for some constant $c>0$, $-\alpha := \max_{\lambda\in\spec(\mathbf{H})} \Re[\lambda]+ \epsilon _H$ for some  $ \epsilon_H >0$ small enough.
{Set $\mathbf e(t):=\mathbf y(t)-\mathbf y^*(t)$. By the fundamental theorem of calculus and $\nabla_{\mathbf y}f_0(\mathbf y^*(t),t)=0$,
\begin{align*}
\nabla_{\mathbf y}f_0(\mathbf y(t),t)&=\mathbf M(t)\mathbf e(t),\\
\mathbf M(t)&:=\int_0^1\nabla_{\mathbf{yy}}f_0(\mathbf y^*(t)+\theta\mathbf e(t),t)\,\mathrm d\theta.
\end{align*}
Assumption~\ref{ass:convexity} gives $\mathbf M(t)\succeq m_f\mathbf I$, and hence
\begin{align*}
\|\mathbf y(t)-\mathbf y^*(t)\|_2\leq m_f^{-1}\|\nabla_{\mathbf y}f_0(\mathbf y(t),t)\|_2.
\end{align*}}

{Combining this bound with \eqref{eq:Left_Norm_Inequality} gives}
\begin{align*}
     &\norm{{\mathbf{y}(t)-\mathbf{y}^*(t)}}_2  \leq C e^{-\alpha t}, \nonumber\\
    & 0 \leq C = \left(\tfrac{c^2}{m_f^2} \sum_{j=0}^{k-1}\nolimits \norm{{\nabla_\mathbf{y}^{(j)} f_0(\mathbf{y}(0),0)}}_2^2)\right)^{\frac{1}{2}} < \infty.
\end{align*}\\
On the other hand, convexity of $f_0(\mathbf{y},t)$ implies that for each $t \geq 0$
\begin{align*}
    0 \!\leq\! f_0(\mathbf{y}(t),t)\! - \!f_0(\mathbf{y}^*(t),t) \!\leq\! \nabla _\mathbf{y} f_0(\mathbf{y}(t),t)^{\T} (\mathbf{y}(t)\!-\!\mathbf{y}^*(t))
\end{align*} By applying Cauchy--Schwarz inequality on the right-hand side we obtain;
\begin{align*}
 0 \leq {f_0(\mathbf{y}(t),t) - f_0(\mathbf{y}^*(t),t)} \leq m_fC^2e^{-2\alpha t}
\end{align*}
which completes the proof.

\subsection{Proof of Lemma \ref{cor: uniform boundedness of inverse KKT}}\label{proof: cor: uniform boundedness of inverse KKT}
For all $t \geq 0$, let $\mu_1(t) \geq \mu_2(t)\geq \cdots \geq {\mu_m(t)} > 0$ be the eigenvalues of  $\nabla_{\mathbf{yy}} f_0(\mathbf{y},t)$. Given Assumption \ref{ass:convexity} and \ref{ass:lipschitz_gradient}, for all $t \geq 0$, we have  ${\mu_m(t)}  \geq m_f > 0$ and  $\mu_1(t)  \leq {L_f} < \infty$. Given Assumption \ref{ass:MFCQ}, we have $\sigma_{\rm min} (\mathbf{A}(t)) \geq \tau_{\rm min}$ and $\sigma_{\rm max} (\mathbf{A}(t)) \leq \tau_{\rm max}$ for all $t \geq 0$.

According to Lemma \ref{lem:Eigenvalues of KKT matrix}, for the positive eigenvalues of $\nabla_{\mathbf{zz}}{L}(\mathbf{z},t)$, we have $m_f$ being their uniform lower bound. Also, for the negative eigenvalues of $\nabla_{\mathbf{zz}}{L}(\mathbf{z},t)$, we have ${\frac{1}{2}(L_f - \sqrt{L_f^2 + 4\tau_{\rm min}^{\,2}})}$ being their uniform upper bound. Consequently, for all $t \geq 0$, we have for all eigenvalues of the KKT matrix
\begin{align}
     \left|\lambda_i\left(\nabla_{\mathbf{zz}}{L}(\mathbf{z},t)\right)\right|  \geq {\min\{m_f, \tfrac{1}{2}(\sqrt{L_f^2 + 4\tau_{\rm min}^{\,2}}-L_f)\}=K},
\end{align}
which leads to
$$ \| \nabla^{-1}_{\mathbf{zz}}{L}(\mathbf{z},t) \|_2 \leq {K^{-1}}.$$

\subsection{Proof of Theorem \ref{Thm: Eq}} \label{Appendix: Eq}
The structure of proof is similar to the proof of Theorem \ref{Thm: Unconstrained}.
According to Lemma \ref{lem:Diffrentiating_Gradient_K_Times}, the trajectory $ \mathbf{z}(t)$ of system \eqref{eq:equality z_kth_lagrange_derivative_solution} satisfy the optimization dynamics as in \eqref{eq:K-th_order_equality_constraint_Optimality_Error}, with $ \mathbf{H}$ being the designed Hurwitz matrix. Similarly, the solution to this ODE satisfies the following inequality:  
\begin{align*}
\sum_{j=0}^{k-1} \norm{{\nabla_\mathbf{z}^{(j)} L(\mathbf{z}(t),t)}}_2^2
    \leq c^2e^{-2\alpha t}\sum_{j=0}^{k-1} \norm{{\nabla_\mathbf{z}^{(j)} L(\mathbf{z}(0),0)}}_2^2
\end{align*}
for some constant $c>0$, $-\alpha := \max_{\lambda\in\spec(\mathbf{H})} \Re[\lambda]+ \epsilon_H $ for some  $ \epsilon_H >0$ small enough.
{Set $\mathbf e_z(t):=\mathbf z(t)-\mathbf z^*(t)$. Since $\nabla_{\mathbf z}L(\mathbf z^*(t),t)=0$, the fundamental theorem of calculus gives
\begin{align*}
\nabla_{\mathbf z}L(\mathbf z(t),t)&=\mathbf M(t)\mathbf e_z(t),\\
\mathbf M(t)&:=\int_0^1\nabla_{\mathbf{zz}}L(\mathbf z^*(t)+\theta\mathbf e_z(t),t)\,\mathrm d\theta.
\end{align*}
Because $\mathbf A(t)$ is independent of $\mathbf z$, this average retains the KKT structure
\begin{gather*}
\mathbf M(t)=\begin{bmatrix}\overline{\mathbf Q}(t)&\mathbf A(t)^{\T}\\\mathbf A(t)&\mathbf 0\end{bmatrix},\quad
m_f\mathbf I\preceq\overline{\mathbf Q}(t)\preceq L_f\mathbf I,\\
\overline{\mathbf Q}(t):={\textstyle\int_0^1}\nabla_{\mathbf{yy}}f_0\bigl((1-\theta)\mathbf y^*(t)+\theta\mathbf y(t),t\bigr)\,\mathrm d\theta.
\end{gather*}
The spectral bound of Lemma~\ref{lem:Eigenvalues of KKT matrix} therefore gives $\|\mathbf M(t)^{-1}\|_2\leq K^{-1}$, with the same $K$ as in Lemma~\ref{cor: uniform boundedness of inverse KKT}. Consequently,
\begin{align*}
\|\mathbf z(t)-\mathbf z^*(t)\|_2\leq K^{-1}\|\nabla_{\mathbf z}L(\mathbf z(t),t)\|_2.
\end{align*}}

{Combining this bound with the preceding gradient estimate gives}
\begin{align*}
    & \| \mathbf{z}(t)-{\mathbf z^*(t)}\|_2 \leq Ce^{-\alpha t} ,\\
    & 0 < C = \left(\tfrac{c^2}{K^2} \sum_{j=0}^{k-1}\nolimits \|{\nabla_\mathbf{z}^{(j)} L(\mathbf{z}(0),0)}\|_2^2)\right)^{\frac{1}{2}} < \infty.\nonumber
\end{align*}

{\subsection{Proof of Proposition \ref{prop:Phi properties}}\label{appendix:relaxed barrier}
\noindent {Fix $\delta>0$.} On the branch $\sigma < \delta$, where $B_\delta$ in \eqref{eq:relaxed barrier} is the degree-$m_r$ polynomial, write $v := (\delta-\sigma)/\delta > 0$. Differentiating term by term with respect to $\sigma$, using $\mathrm{d}v/\mathrm{d}\sigma = -1/\delta$, gives
$B_\delta'(\sigma) = -\tfrac{1}{\delta}\sum_{i=0}^{m_r-1} v^{i}$ and $B_\delta''(\sigma) = \tfrac{1}{\delta^2}\sum_{j=2}^{m_r}(j-1)v^{j-2}$. In each sum the first term equals $1$ and the remaining ones are nonnegative, so $B_\delta'(\sigma) \le -\tfrac1\delta$ and $B_\delta''(\sigma) \ge \tfrac{1}{\delta^2} > 0$. On the branch $\sigma \ge \delta$, $B_\delta = -\log$ is convex and strictly decreasing, and the derivatives of the two branches match up to order $m_r$ at $\sigma = \delta$ by the Taylor construction. Hence $B_\delta \in C^{m_r}(\real)$, convex and strictly decreasing. Since each $f_i(\cdot,t)$ is convex (Assumption \ref{ass:convexity}), $\sigma_i(\mathbf{y},t) := s(t) - f_i(\mathbf{y},t)$ is concave in $\mathbf{y}$, and $B_\delta \circ \sigma_i$, a nonincreasing convex function of a concave function, is convex \cite[Sec.~3.2.4]{boyd2004convex}. Adding the $m_f$-strongly convex $f_0$ makes $\Phi_\delta(\cdot,t)$ $m_f$-strongly convex and $C^{m_r}$ on $\real^m$, whence $\nabla_{\mathbf{yy}}\Phi_\delta \succeq m_f \mathbf{I}$ and $\|\nabla^{-1}_{\mathbf{yy}}\Phi_\delta\| \le 1/m_f$ globally. Existence and uniqueness of $\hat{\mathbf{y}}^*_\delta(t)$ follow from strong convexity, which proves (i) and the first claim of (ii).
{For the radius bound, let $0<\delta(t)<\rho$. The point $\mathbf y_{\rm s}(t)$ lies in the logarithmic branch, since $s(t)-f_i(\mathbf y_{\rm s}(t),t)\geq\rho$. Thus
\begin{align*}
\|\nabla_{\mathbf y}\Phi_\delta(\mathbf y_{\rm s}(t),t)\|_2
&\leq M_0(R_{\rm s})+\frac{\sum_{i=1}^p M_i(R_{\rm s})}{c_0\rho}=K_{\rm s},
\end{align*}
and strong convexity gives $\|\hat{\mathbf y}^*_\delta(t)-\mathbf y_{\rm s}(t)\|_2\leq K_{\rm s}/m_f$. Likewise, $f_0(\mathbf y^*(t),t)\leq f_0(\mathbf y_{\rm s}(t),t)$ and strong convexity give $\|\mathbf y^*(t)-\mathbf y_{\rm s}(t)\|_2\leq2M_0(R_{\rm s})/m_f$. Hence both minimizers lie in the ball of radius $R$, independently of the schedules $c(t)\geq c_0$, $s(t)\geq0$, and $\delta(t)$ above.}
}

{\subsection{Proof of Lemma \ref{lem:relaxed barrier}(i)}\label{appendix:exactness}
\noindent {Let $0<\delta(t)<\rho$. Set $\bar\sigma=\epsilon_{\rm r}:=\rho/2$ and $\mathbb I_{\bar\sigma}(t):=\{i\in[p]:s(t)-f_i(\hat{\mathbf y}^*_\delta(t),t)\leq\bar\sigma\}$. By Proposition~\ref{prop:Phi properties}(ii), the direction $\mathbf d_\delta(t):=\mathbf y_{\rm s}(t)-\hat{\mathbf y}^*_\delta(t)$ satisfies $\|\mathbf d_\delta(t)\|_2\leq R_{\rm s}+R=:d_{\rm r}$. For each $i\in\mathbb I_{\bar\sigma}(t)$, convexity and Assumption~\ref{ass:lipshictz_continuity} give
\begin{align*}
\nabla_{\mathbf y}f_i(\hat{\mathbf y}^*_\delta(t),t)^{\T}\mathbf d_\delta(t)
&\leq f_i(\mathbf y_{\rm s}(t),t)-f_i(\hat{\mathbf y}^*_\delta(t),t)\\
&\leq-\rho-s(t)+\bar\sigma\leq-\epsilon_{\rm r}.
\end{align*}
Thus $d_{\rm r},\epsilon_{\rm r},\bar\sigma$ are independent of these schedules.

For (i), assume $0<\delta(t)<\min\{\rho,1/(\Lambda c(t))\}$. The definition~\eqref{eq:Lambda} gives $\Lambda\geq1/(c_0\bar\sigma)$, so $\delta(t)<\bar\sigma$.} At the minimizer, stationarity of $\Phi_\delta(\cdot,t)$ gives
\begin{equation*}
\nabla_{\mathbf{y}} f_0(\hat{\mathbf{y}}^*_\delta(t),t) + \sum_{i=1}^p \tilde\lambda_i(t)\, \nabla_{\mathbf{y}} f_i(\hat{\mathbf{y}}^*_\delta(t),t) = 0,
\end{equation*}
with $\tilde\lambda_i(t) := -B_{\delta(t)}'(\sigma_i(t))/c(t) > 0$ and $\sigma_i(t) := s(t) - f_i(\hat{\mathbf{y}}^*_\delta(t),t)$. {We split the indices at $\bar\sigma=\rho/2$.} For $\sigma_i(t) \ge \bar\sigma$ (log branch),
\begin{equation*}
\tilde\lambda_i(t) = \frac{1}{c(t)\,\sigma_i(t)} \le \frac{1}{c_0\bar\sigma}.
\end{equation*}
{Taking the inner product of stationarity with $\mathbf d_\delta(t)$ and using the directional bound above yields
\begin{align*}
\epsilon_{\rm r} \!\!\sum_{i \in \mathbb{I}_{\bar\sigma}(t)}\!\! \tilde\lambda_i(t)
&\le \Big(\nabla_{\mathbf{y}} f_0 + \!\!\sum_{i \notin \mathbb{I}_{\bar\sigma}(t)}\!\! \tilde\lambda_i(t)\, \nabla_{\mathbf{y}} f_i\Big)^{\!\T} \mathbf d_\delta(t) \\
&\le d_{\rm r}\left(M_0(R)+\frac{\sum_{i=1}^p M_i(R)}{c_0\bar\sigma}\right),
\end{align*}}

{where the last step uses the bounds $M_i(R)$ of Assumption~\ref{ass:lipshictz_continuity}, $\|\mathbf d_\delta(t)\|_2\leq d_{\rm r}$, and the log-branch bound above (arguments $(\hat{\mathbf y}^*_\delta(t),t)$ omitted).}

Hence $\tilde\lambda_i(t) \le \Lambda$ for all $i$ and all $t \ge 0$, with
\begin{equation}\label{eq:Lambda}
{\Lambda:=\max\left\{\frac{d_{\rm r}}{\epsilon_{\rm r}}\left(M_0(R)+\frac{\sum_{i=1}^p M_i(R)}{c_0\bar\sigma}\right),\frac{1}{c_0\bar\sigma}\right\}.}
\end{equation}
Now suppose $\sigma_i(t) < \delta(t)$ for some $i$ and $t$; by Appendix \ref{appendix:relaxed barrier}, $|B_{\delta(t)}'(\sigma_i(t))| \ge 1/\delta(t)$, so $\tilde\lambda_i(t) \ge 1/(c(t)\delta(t)) > \Lambda$ whenever $\delta(t) < 1/(\Lambda c(t))$, a contradiction. {Thus $\sigma_i(t)\geq\delta(t)$ for every $i$, so each barrier term and its derivative agree with the logarithmic barrier at $\hat{\mathbf{y}}^*_\delta(t)$. Hence $\nabla_{\mathbf{y}}\Phi_0(\hat{\mathbf{y}}^*_\delta(t),t)=0$, and strict convexity gives $\hat{\mathbf{y}}^*_\delta(t)=\hat{\mathbf{y}}^*_0(t)$.}   \hfill$\blacksquare$
}

\subsection{Proof of Lemma \ref{lem:Unifrom_boundedness_Lagrange_Constraints}} \label{Appendix: Uniform_Boundedness_Multiplier}
{By Assumption~\ref{ass:lipshictz_continuity}, the problem has a strictly feasible point. Convexity therefore ensures existence of KKT multipliers. For any such $\boldsymbol\lambda^*(t)$, complementarity makes the inactive multipliers zero. Taking the inner product of stationarity with the MFCQ direction $\Bar{\mathbf d}(t)$ gives
\begin{align*}
\epsilon\|\boldsymbol\lambda^*(t)\|_1
&\leq-\sum_{i\in\mathbb I(\mathbf y^*(t))}\lambda_i^*(t)
\nabla_{\mathbf y}f_i(\mathbf y^*(t),t)^{\T}\Bar{\mathbf d}(t)\\
&=\nabla_{\mathbf y}f_0(\mathbf y^*(t),t)^{\T}\Bar{\mathbf d}(t)
\leq M_0(R)d.
\end{align*}
Here $\|\mathbf y^*(t)\|_2\leq R$ by Proposition~\ref{prop:Phi properties}(ii). Dividing by $\epsilon$ proves the bound for every KKT multiplier.}

\subsection{Proof of Theorem \ref{Thm: Ineq}} \label{Appendix: Ineq}
{\noindent Fix an admissible $\delta(t)>0$ and write $\hat{\mathbf{y}}^*(t):=\hat{\mathbf{y}}^*_\delta(t)$.}

The structure of proof is similar to the proof of Theorem \ref{Thm: Unconstrained}.
According to Lemma \ref{lem:Diffrentiating_Gradient_K_Times}, the trajectory $ \mathbf{y}(t)$ of system \eqref{eq:ineq_design_reqirement} satisfy the optimization dynamics as in \eqref{eq:barrier_function_Dynamical_System}, with $ \mathbf{H}$ being the designed Hurwitz matrix. Similarly, the solution to this ODE satisfies the following inequality:  
\begin{align*}
\sum_{j=0}^{k-1} \norm{{\nabla_\mathbf{y}^{(j)} {\Phi_\delta}(\mathbf{y}(t),t)}}_2^2
    \leq c^2e^{-2\alpha t}\sum_{j=0}^{k-1} \norm{{\nabla_\mathbf{y}^{(j)} {\Phi_\delta}(\mathbf{y}(0),0)}}_2^2
\end{align*}
for some constant $c>0$, $-\alpha := \max_{\lambda\in\spec(\mathbf{H})} \Re[\lambda]+ \epsilon_H $ for some  $ \epsilon_H >0$ small enough.

{Set $\mathbf e(t):=\mathbf y(t)-\hat{\mathbf y}^*(t)$. The fundamental theorem of calculus and stationarity at $\hat{\mathbf y}^*(t)$ give
\begin{align*}
\nabla_{\mathbf y}\Phi_\delta(\mathbf y(t),t)&=\mathbf M(t)\mathbf e(t),\\
\mathbf M(t)&:=\int_0^1\nabla_{\mathbf{yy}}\Phi_\delta(\hat{\mathbf y}^*(t)+\theta\mathbf e(t),t)\,\mathrm d\theta.
\end{align*}
By Proposition~\ref{prop:Phi properties}(i), $\mathbf M(t)\succeq m_f\mathbf I$. Hence
\begin{align}
\|\mathbf y(t)-\hat{\mathbf y}^*(t)\|_2\leq m_f^{-1}\|\nabla_{\mathbf y}\Phi_\delta(\mathbf y(t),t)\|_2.\label{eq: mean_value_thm_Ineq}
\end{align}}

{Combining \eqref{eq: mean_value_thm_Ineq} with the preceding gradient estimate gives}
\begin{align*}
     &\norm{{\mathbf{y}(t)-\mathbf{\hat{y}}^*(t)}}_2  \leq C e^{-\alpha t}, \nonumber\\
    & 0 \leq C = \left(\tfrac{c^2}{m_f^2} \sum_{j=0}^{k-1}\nolimits \norm{{\nabla_\mathbf{y}^{(j)}  {\Phi_\delta}(\mathbf{y}(0),0)}}_2^2)\right)^{\frac{1}{2}} < \infty.
\end{align*}
{Proposition~\ref{prop:Phi properties}(ii) and the tracking estimate above give $\|\mathbf y(t)\|_2\leq R+C$. The segment joining $\hat{\mathbf y}^*(t)$ to $\mathbf y(t)$ therefore lies in the ball of radius $R+C$. Applying the fundamental theorem of calculus along this segment and the gradient bound of Assumption~\ref{ass:lipshictz_continuity} gives
\begin{align}
|f_0(\mathbf y(t),t)-f_0(\hat{\mathbf y}^*(t),t)|
\leq M_0(R+C)Ce^{-\alpha t}.\label{eq:Proof_Error_y_yhatstar}
\end{align}
Lemmas~\ref{lem:sub-optimality_gap_perturbed_approximation} and~\ref{lem:Unifrom_boundedness_Lagrange_Constraints} give
\begin{equation}
|f_0\!(\hat{\mathbf y}^*\!(t),\!t)\!-\!f_0\!(\mathbf y^*\!(t),\!t)|\!\leq\!\frac p{c_0}e^{-\alpha_c t}\!+\!\frac{M_0\!(R)d}{\epsilon}s_0e^{-\alpha_s t}.\label{eq:Proof_error_yhatstar_ystar}
\end{equation}
By the triangle inequality, the objective-error bound follows with $\beta:=\min\{\alpha,\alpha_c,\alpha_s\}$,
\begin{align*}
C'&:=M_0(R+C)C+\frac p{c_0}+\frac{M_0(R)d s_0}{\epsilon}.
\end{align*}
The prefactor $C'$ may depend on the initial condition, while $\beta$ does not. Finally, Lemma~\ref{lem:relaxed barrier}(i) gives $f_i(\hat{\mathbf y}^*(t),t)<s(t)$. Strong convexity of $L(\cdot,\boldsymbol\lambda^*(t),t)$, KKT stationarity, and Lemma~\ref{lem:relaxed barrier}(ii) then imply
\begin{align*}
\frac{m_f}{2}\|\hat{\mathbf y}^*(t)-\mathbf y^*(t)\|_2^2
\leq\frac p{c(t)}+\frac{2M_0(R)d}{\epsilon}s(t)\to0.
\end{align*}}

\bibliographystyle{IEEEtran}
\bibliography{main}

\begin{IEEEbiography}
[{\includegraphics[width=1in,height=1.25in,clip,keepaspectratio]{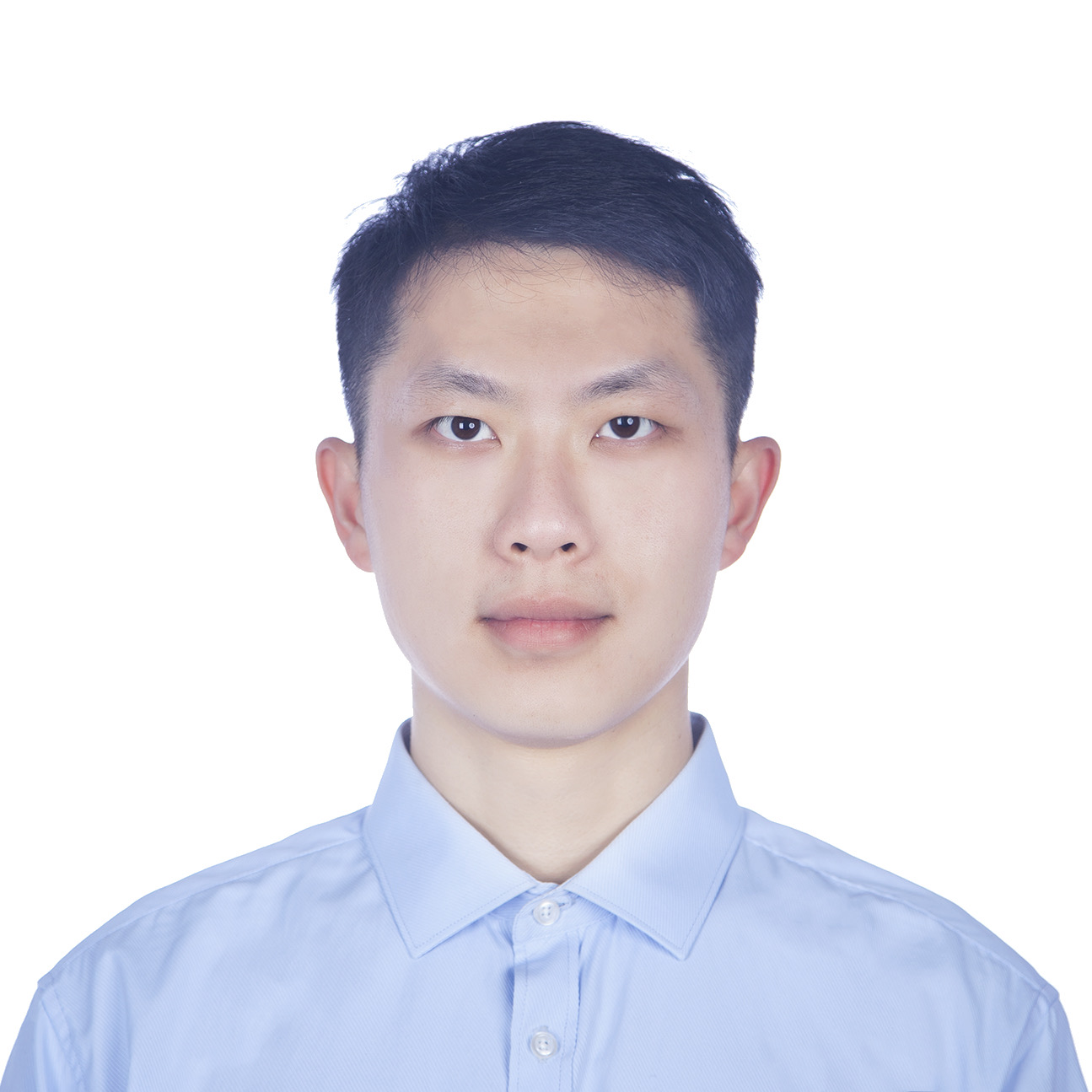}}]{Tianqi Zheng}
is currently working toward his Ph.D. degree at the Department of Electrical and Computer Engineering, Johns Hopkins University. He received his B.Eng. degree in Electrical Engineering and Automation from the Harbin Institute of Technology in 2018, and his M.S.E degree in Applied Mathematics and Statistics from Johns Hopkins University in 2023. His research interests include control theory, optimization and reinforcement learning.
\end{IEEEbiography}
\begin{IEEEbiography}[{\includegraphics[width=1in,height=1.25in,clip,keepaspectratio]{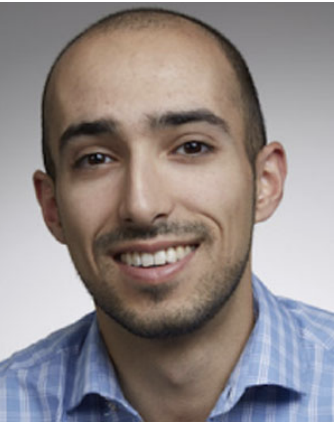}}]{John W. Simpson-Porco} (S'10--M'15--SM'22--) received the B.Sc. degree in engineering physics from Queen's University, Kingston, ON, Canada in 2010, and the Ph.D. degree in mechanical engineering from the University of California at Santa Barbara, Santa Barbara, CA, USA in 2015. He is currently an Associate Professor of Electrical and Computer Engineering at the University of Toronto, Toronto, ON, Canada. He was previously an Assistant Professor at the University of Waterloo, Waterloo, ON, Canada and a visiting scientist with the Automatic Control Laboratory at ETH Z\"{u}rich, Z\"{u}rich, Switzerland. His research focuses on feedback control theory and applications of control in modernized power grids. He is a recipient of the Automatica Paper Prize, the Center for Control, Dynamical Systems and Computation Best Thesis Award, the IEEE PES Technical Committee Working Group Recognition Award for Outstanding Technical Report, and the Ontario Early Researcher Award.
\end{IEEEbiography}

\begin{IEEEbiography}[{\includegraphics[width=1in,height=1.25in,clip,keepaspectratio]{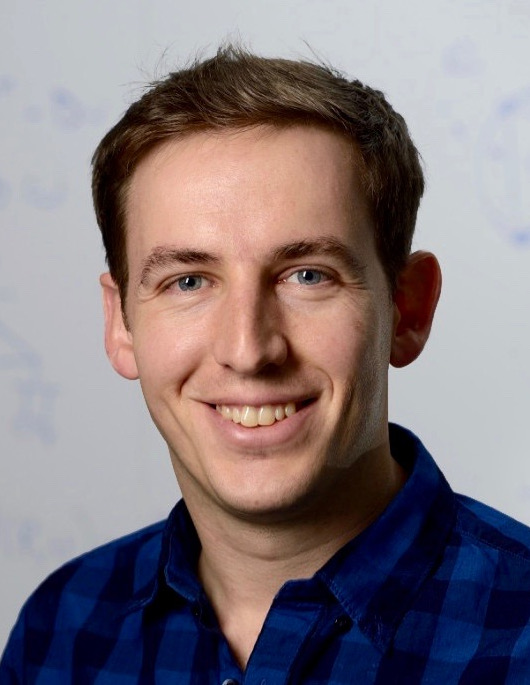}}]
{Enrique Mallada} (S'09--M'13--SM'19--) is an Associate Professor of Electrical and Computer Engineering at Johns Hopkins University since 2022. Prior to joining Hopkins in 2016, he was a Post-Doctoral Fellow in the Center for the Mathematics of Information at Caltech from 2014 to 2016. He received his Ingeniero en Telecomunicaciones degree from Universidad ORT, Uruguay, in 2005 and his Ph.D. degree in Electrical and Computer Engineering with a minor in Applied Mathematics from Cornell University in 2014.
Dr. Mallada was awarded
the Johns Hopkins Alumni Association Teaching Award in 2021,
the Catalyst and Discovery Awards in 2020 and 2021, respectively, from Johns Hopkins University,
the NSF CAREER award in 2018,
the ECE Director's Ph.D. Thesis Research Award for his dissertation in 2014,
the Center for the Mathematics of Information (CMI) Fellowship from Caltech in 2014,
and the Cornell University Jacobs Fellowship in 2011.
His research interests lie in the areas of control, dynamical systems, and optimization, with applications to engineering networks.
\end{IEEEbiography}

\end{document}